\documentclass[aps,prx,reprint,preprintnumbers,superscriptaddress,nofootinbib,longbibliography,floatfix]{revtex4-2}

\usepackage[utf8]{inputenc}
\usepackage[T1]{fontenc}
\usepackage{lmodern}
\usepackage{microtype}
\usepackage{mathtools}

\usepackage{placeins}

\usepackage{amsmath}

\usepackage{amsmath}
\usepackage[caption=false]{subfig}
\usepackage{amsfonts}
\usepackage{graphicx}
\usepackage{xspace}
\usepackage{hyperref}
\hypersetup{
  colorlinks=true,
  citecolor=blue,
  linkcolor=blue,
  urlcolor=blue
}

\usepackage{color}
\definecolor{darkred}{rgb}{1.0,0.1,0.1}
\definecolor{darkgreen}{rgb}{0.1,0.7,0.1}
\definecolor{darkblue}{rgb}{0.1,0.1,1.0}

\newcommand{\omnifold}{\textsc{OmniFold}\xspace}
\newcommand{\multifold}{\textsc{MultiFold}\xspace}
\newcommand{\Delphes}{\textsc{Delphes}\xspace}
\newcommand{\pt}{$p_{\mathrm{T}}$}

\begin{document}

\preprint{MIT-CTP 5305}

\title{Preserving New Physics while Simultaneously Unfolding All Observables}

\author{Patrick Komiske}
\email{pkomiske@mit.edu}
\affiliation{Center for Theoretical Physics, Massachusetts Institute of Technology, Cambridge, MA 02139, USA}
\affiliation{The NSF AI Institute for Artificial Intelligence and Fundamental Interactions}

\author{W. Patrick McCormack}
\email{wpmccormack@lbl.gov}
\affiliation{Department of Physics, University of California, Berkeley, CA 94720, USA}
\affiliation{Physics Division, Lawrence Berkeley National Laboratory, Berkeley, CA 94720, USA}

\author{Benjamin Nachman}
\email{bpnachman@lbl.gov}
\affiliation{Physics Division, Lawrence Berkeley National Laboratory, Berkeley, CA 94720, USA}
\affiliation{Berkeley Institute for Data Science, University of California, Berkeley, CA 94720, USA}

\begin{abstract}
Direct searches for new particles at colliders have traditionally been factorized into model proposals by theorists and model testing by experimentalists.  With the recent advent of machine learning methods that allow for the simultaneous unfolding of all observables in a given phase space region, there is a new opportunity to blur these traditional boundaries by performing searches on unfolded data.  This could facilitate a research program where data are explored in their natural high dimensionality with as little model bias as possible.  We study how the information about physics beyond the Standard Model is preserved by full phase space unfolding using an important physics target at the Large Hadron Collider (LHC): exotic Higgs boson decays involving hadronic final states.  We find that if the signal cross section is high enough, information about the new physics is visible in the unfolded data.  We will show that in some cases, quantifiably all of the information about the new physics is encoded in the unfolded data.  Finally, we show that there are still many cases when the unfolding does not work fully or precisely, such as when the signal cross section is small.  This study will serve as an important benchmark for enhancing unfolding methods for the LHC and beyond.
\end{abstract}

\maketitle


\tableofcontents

\section{Introduction}
\label{sec:intro}

Analyses at the Large Hadron Collider (LHC) are generally classified as \textit{measurements} or \textit{searches} if their goal is to search for indirect or direct signs of physics beyond the Standard Model (SM), respectively.  An important reason for this distinction is that measurements assume that deviations to the SM are small. This is required so that the removal of detector distortions (`unfolding') can be based on SM simulations.  Traditional unfolding methods~\cite{Cowan:2002in,Blobel:2203257,doi:10.1002/9783527653416.ch6,Balasubramanian:2019itp,DAgostini:1994fjx,Hocker:1995kb,Schmitt:2012kp} are based on low-dimensional and binned observables.  The detector response may depend on additional unmeasured features and may vary strongly within a given bin.  If these properties are significantly different for new particles, then an unfolding derived with SM simulations is likely to be inaccurate.  

This feature of current unfolding methods has been studied in~\cite{Facini:2019rgg} and limits the applicability of recasting tools such as \textsc{Contur}~\cite{Butterworth:2016sqg}. A variety of complementary tools have been developed to fold model predictions with a detector response, including \textsc{MadAnalysis}~\cite{Conte:2012fm,Conte:2014zja,Conte:2018vmg,Dumont:2014tja,Araz:2019otb}, \textsc{Recast}~\cite{Cranmer:2010hk,ATLAS:2020viz,ATLAS:2019ivx}, \textsc{CheckMATE}~\cite{Drees:2013wra,Dercks:2016npn}, \textsc{SModelS}~\cite{Kraml:2013mwa,Alguero:2020buz}, \textsc{Fastlim}~\cite{Papucci:2014rja}, and~\textsc{XQCAT}~\cite{Barducci:2014ila}. In addition to limitations from recasting approximations, these approaches are limited by the minimal (binned) search results that are usually highly optimized for particular signal models.

One possibility is to perform model-agnostic approaches at detector-level using one of the growing number of anomaly detection methods~\cite{Collins:2018epr,DAgnolo:2018cun,Collins:2019jip,DAgnolo:2019vbw,Farina:2018fyg,Heimel:2018mkt,Roy:2019jae,Cerri:2018anq,Blance:2019ibf,Hajer:2018kqm,DeSimone:2018efk,Mullin:2019mmh,1809.02977,Dillon:2019cqt,Andreassen:2020nkr,Nachman:2020lpy,Aguilar-Saavedra:2017rzt,Romao:2019dvs,Romao:2020ojy,knapp2020adversarially,collaboration2020dijet,1797846,1800445,Amram:2020ykb,Cheng:2020dal,Khosa:2020qrz,Thaprasop:2020mzp,Alexander:2020mbx,aguilarsaavedra2020mass,1815227,pol2020anomaly,Mikuni:2020qds,vanBeekveld:2020txa,Park:2020pak,Faroughy:2020gas,Stein:2020rou,Kasieczka:2021xcg,Chakravarti:2021svb,Batson:2021agz,Blance:2021gcs,Bortolato:2021zic,Collins:2021nxn} (for reviews, see Ref~\cite{Nachman:2020ccu,Kasieczka:2021xcg}).  These techniques can achieve broad and deep sensitivity by learning directly from data.  However, methods that do not rely on any signal information (unsupervised) are not particularly sensitive~\cite{Amram:2020ykb,Collins:2021nxn} and methods that use noisy or partial signal information (weakly and semisupervised, respectively) are not recastable after the search is performed~\cite{collaboration2020dijet}.

A new solution that has emerged is to perform an unbinned unfolding using all of the available information.  If the high frequency and high dimensional aspects of the detector response are part of the unfolding procedure, then differences between signal and background will not be a source of bias.  Unbinned and high-dimensional unfolding are now possible with advances in machine learning~\cite{Andreassen:2019cjw,Bellagente:2020piv,Bellagente:2019uyp} (for other machine learning and unbinned proposals, see Ref.~\cite{Aslan:2003vu,Lindemann:1995ut,Gagunashvili:2010zw,Glazov:2017vni,Datta:2018mwd}).  Of these, only the \textsc{OmniFold}~\cite{Andreassen:2019cjw} can currently process the full phase space that includes all observable particles and their properties.  Unlike proposals based on generative models, \textsc{OmniFold} is built on neural network classifiers that are used to iteratively reweight simulations to match the data.  Classifiers designed to process variable-length, unordered sets of particles allow this technique to access the full phase space~\cite{10.5555/3294996.3295098,Komiske:2018cqr}.  While \textsc{OmniFold} has yet to be applied to collider data in the full phase space, it has recently been deployed in a low-dimensional case with the H1 experiment~\cite{H1prelim-21-031}.

In this paper, we investigate the ability of \textsc{OmniFold} to preserve information about new particles present in the data.  In particular, we will study the direct production of new particles which have spectra that are not similar to the SM background.  Our benchmark example will be the exotic decay of a Higgs boson-like particle decaying into a $Z$ boson and a light color singlet that decays into hadrons.  The dominant background to this process is the SM production of $Z$ bosons and jets.  We will see to what extent the information about new particles are presevered in the unfolding.  Recently, the authors of Ref.~\cite{Bellagente:2019uyp} showed that generative model approaches can preserve new physics with a relatively large cross section.  Our first example will be motivated by this example and then we will explore how the sensitivity depends on variations in signal model parameters and unfolding setup. 

This paper is organized as follows.  Section~\ref{sec:methods} briefly reviews full phase space unfolding and introduces the benefits and challenges of the existing approach in the context of physics beyond the Standard Model.  The simulation samples and machine learning setup are introduced in Sec.~\ref{sec:setup}.  Section~\ref{sec:Higgs250} explores a case where a model-independent new physics search technique, such as bump hunting, could be applied in unfolded data.  Section~\ref{sec:results} then studies an example of exotic Higgs boson decays, where simple bump-hunting would be less fruitful.  Implications of model-dependent search program in unfolded data are explored in this section as well.  The paper ends with conclusions and outlook in Sec.~\ref{sec:conclusions}.

\section{Review of \omnifold Unfolding}
\label{sec:methods}

\begin{figure}
  \centering
  \includegraphics[width=0.45\textwidth]{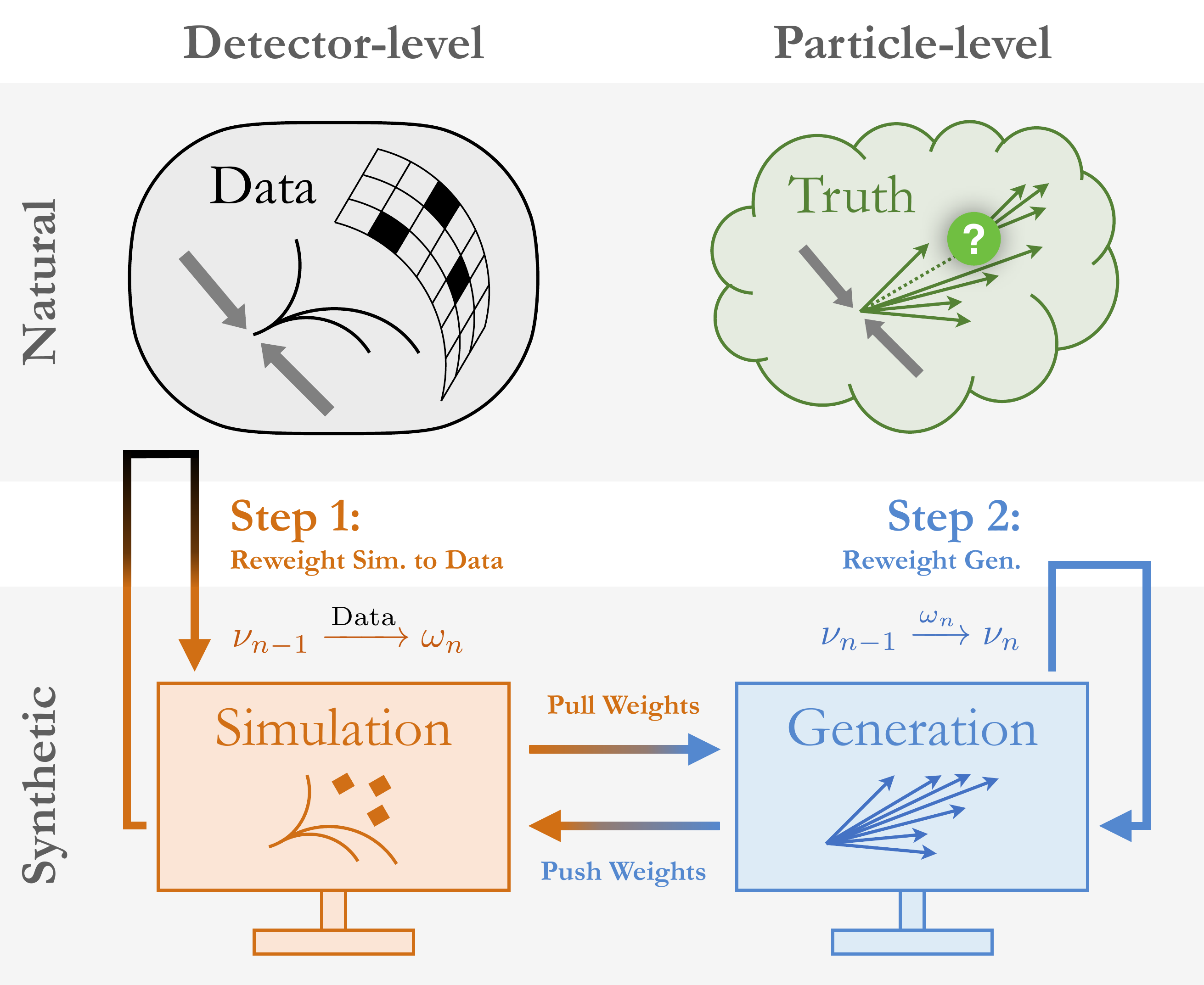}
  \caption{Visual representation of the \textsc{OmniFold} method.  This method relies on experimental ``Data'', which was caused by some underlying ``Truth'' distribution, and synthetic datasets at both particle-level (``Generation'') and detector-level (``Simulation'').  The Simulation is reweighted to match the Data using a classifer, and these weights, $\omega_i$ are applied to Generation.  The unweighted Generation is reweighted to match the Generation with $\omega_i$ applied, giving particle-level weights, $\nu_i$.  This process is repeated iteratively, as the $\nu_i$ are applied to the Simulation, and this reweighted Simulation is once again reweighted to Data.  Based on the corresponding figure from Ref.~\cite{Andreassen:2019cjw}.}
  \label{fig:omnifold}
\end{figure}

The \textsc{OmniFold} method is represented visually in Fig.~\ref{fig:omnifold}.  There are two inputs: natural data from experiment and synthetic data from simulation.  The goal is to remove the detector distortions from the observations (``Data'') to infer the underlying particle-level distribution (``Truth'').  Synthetic particle-level events (``Generation'') provide the initial guess for the Truth and we have an event-by-event match between the Generation and detector-level synthetic data (``Simulation''). As in all unfolding algorithms, we assume that the detector response is well-modeled.

The first step in the \textsc{OmniFold} method is to train a classifier to distinguish the Data from the Simulation.  An optimally trained classifier that minimizes one of the standard loss functions (cross entropy or mean squared error) will predict the probability that an event $x_\text{det}$ is drawn from Data instead of Simulation: $\mathrm{Pr}( \mathrm{Data} | x_\text{det})$.  By applying the weight
\begin{equation}
    \omega_{1} = \frac{\mathrm{Pr}( \mathrm{Data} | x_\text{det})}{1 - \mathrm{Pr}( \mathrm{Data} | x_\text{det})}\propto\frac{p(x_\text{det}|\mathrm{Data})}{p(x_\text{det}|\mathrm{Simulation})}
\end{equation}
to each simulated event, the Simulation will closely resemble the Data.

The detector-level weights, $\omega_{1}$ are then applied to the corresponding particle-level events, resulting in a reweighted distribution of particle-level events.  A second classification step is required because these \textit{pulled back} weights are not a proper function of the particle-level phase space: the same phase space point can be mapped to two different detector-level points with different weights under the stochastic mapping of the detector.  The second classifier distinguishes the nominal Generation from the one using the weights from the first step.  As with the first step, an optimally trained classifier will learn 
\begin{align}
    \nu_{1} &= \frac{\mathrm{Pr}( \mathrm{reweighted~Gen.} | x_\text{part})}{1 - \mathrm{Pr}( \mathrm{reweighted~Gen.} | x_\text{part})}\\
    &\propto \frac{p(x_\text{part}|\mathrm{reweighted~Gen.})}{p(x_\text{part}|\mathrm{Generation})}\,.
\end{align}
The matching between Generation and Simulation can be used to \textit{push} the weights to detector-level and the entire process can be repeated $N$ times.  The final result is the Generation dataset with a set of per-event weights $v_N$.



We will parameterize the classifiers as neural networks.  When $x$ is the full phase space, i.e. a complete list of reconstructed or true particles with their observable properties, we need a neural network architecture that can process variable length, unordered sets.  For this purpose, we use Particle Flow Networks (PFN)~\cite{10.5555/3294996.3295098, Komiske:2018cqr}.   A reduced alternative approach will use a fixed number of high-level observables, which will use a standard fully connected neural network.  To distinguish these two cases, we will call the full phase space version \omnifold and the reduced version \multifold. 


\subsection{\omnifold in the Presence of New Physics}
\label{sec:limitations}

The main benefit of the \omnifold~method comes from its use of low-level observables and the freedom from fixed bins.  By using information about each reconstructed particle in an event or jet, the full phase space is exploited.  Therefore, as long as the interaction of individual particles with the the detector is modelled well, beyond the Standard Model (BSM) physics should not negatively affect the ability to unfold.  If there is BSM physics present in the data, then the most BSM-like events in Simulation will be upweighted as appropriate\footnote{Events may be downweighted in the case of interference effects}.  In traditional unfolding schemes that use regularized matrix inversion of binned histograms, the presence of beyond the Standard Model (BSM) physics could affect the detector response matrix in ways that are not accounted for in a SM-only simulation.  \omnifold~is less affected by this, and is not affected by the possibility of sub-optimal binning choices for BSM sensitivity.

However, there is a key assumption in the \omnifold~method: the initial Simulation and the Data must have overlapping support.   For example, if a heavy resonance existed at a mass well beyond the last data point in Simulation, then it would not be possible to upweight events to match the resonance even in a binned histogram case.  In the \omnifold~case, the initial Simulation should span the data in \textit{all} dimensions.  Empirically, the Simulations often used at the LHC share the same support as Data.  In practice, one needs the ratios of probability densities to not be too far from unity because even if the support is overlapping, a very small likelihood ratio will have a large weight and thus poor statistical uncertainty.


\section{Simulation and Machine Learning Setup}
\label{sec:setup}

Many models of new physics predict new heavy particles that decay to Standard Model particles.  If the invariant mass of the decay particles is computed, a resonant enhancement in the mass spectrum should occur, centered on the new particle's mass.  As an initial exploration of the efficacy of \omnifold~in the presence of BSM physics, we consider the case of a new heavy scalar particle.
Our study is based on proton-proton collisions generated at $\sqrt{s}=14$ TeV with Tune 26~\cite{ATL-PHYS-PUB-2014-021} of \textsc{Pythia}~8.243~\cite{Sjostrand:2007gs,Sjostrand:2006za,Sjostrand:2014zea}.
Signal events are generated as $h\rightarrow Za, a\rightarrow gg$, where $m_h$ has been set to 125 or 250 GeV, and various $a$ masses have been used.  This final state (with low $m_a$ and $m_h=125$ GeV) was recently studied by the ATLAS Collaboration in Ref.~\cite{Aad:2020hzm}.
Detector effects are emulated with \Delphes~3.4.2~\cite{deFavereau:2013fsa}, using the CMS detector card, which uses particle flow reconstruction.
For this study, the Data, Truth, Simulation, and Generation sets consist of 200,000 events. 
Jets with radius parameter $R=0.4$ are clustered using either all particle flow objects (detector-level) or stable non-neutrino truth particles (particle-level) with the anti-$k_T$ algorithm~\cite{Cacciari:2008gp} implemented in \textsc{FastJet}~3.3.2~\cite{Cacciari:2011ma,Cacciari:2005hq}.
We consider leptonic decays of the $Z$ boson, which can be precisely reconstructed.  The target final state is then $Z$ boson production in association with one jet that has non-trivial substructure.  Events are selected if there is at least one truth-level and one detector-level particle within the jet\footnote{We ignore acceptance effects from the jet selection, which can be made arbitrarily small in this case by using only the $Z$ to choose events.}. 
%



For \omnifold, we use all of the particles in the leading jet. Each particle is specified with its \pt, rapidity, $y$, azimuthal angle, $\phi$, and particle identification number.  Furthermore, the invariant mass of the $Z$+jet system, the jet mass, and the jet multiplicity are included as global features.  These data are processed using PFNs implemented in the \textsc{EnergyFlow} Python package~\cite{energyflow}. The PFN architecture is composed of an encoder followed by a fully connected network.  The encoder has two hidden layers of 200 nodes each and outputs a 256-dimensional latent vector.  These vectors are summed over all particles and then the subsequent fully connected network is composed of three layers of 100 nodes each.

For \multifold, we use ten features from each event, based on the $Z$ boson properties and the leading jet.  These features include the invariant mass of the $Z$+jet system, the jet mass, the jet constituent multiplicity, the jet \pt, the $Z$ \pt, the jet Les Houches Angularity~\cite{badger2016les, Gras:2017jty}, the jet width~\cite{Catani:1992jc, 1981NuPhB.191...63R, Ellis:1986ig, Gras:2017jty}, the groomed jet mass with Soft Drop parameters $z_{\mathrm{cut}} = 0.1$ and $\beta = 0$~\cite{Larkoski:2014wba}, the groomed jet momentum fraction (same Soft Drop parameters), and the jet image activity, which is the minimum number of pixels in a jet image that contain 95\% the total \pt~\cite{Pumplin:1991kc}.  The \multifold~neural networks are composed of three hidden layers of 100 nodes each.

For each iteration of \omnifold~and \multifold, the neural network was trained with 120 and 20 epochs, respectively, and included an early stopping condition based on validation loss improvement.  The validation sample was constructed from a random 20\% of the events.  The models are randomly initialized in the first iteration and subsequently warm-started using the model from the previous iteration. All neural networks are implemented using \textsc{Keras}~\cite{keras} with the \textsc{Tensorflow} backend~\cite{tensorflow} and optimized with \textsc{Adam}~\cite{adam}.



\section{Heavy Scalar Decay Study}
\label{sec:Higgs250}

First, we study the case of $m_h=250$ GeV where 10\% of the data is BSM physics.  This composition and signal model relative to the $Z$+jets background are qualitatively similar to the example presented in Ref.~\cite{Bellagente:2019uyp}, which used generative models.

\begin{figure}
  \centering
  \includegraphics[height=3.5cm]{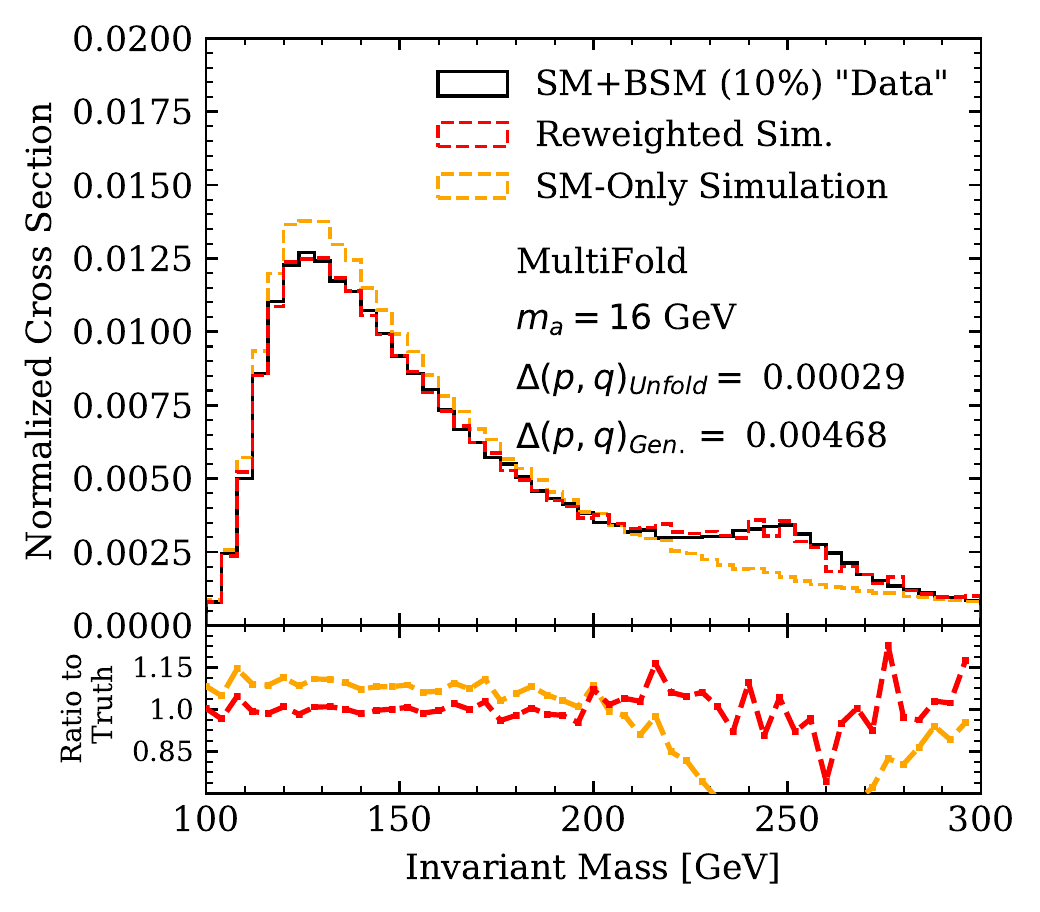}
  \includegraphics[height=3.5cm]{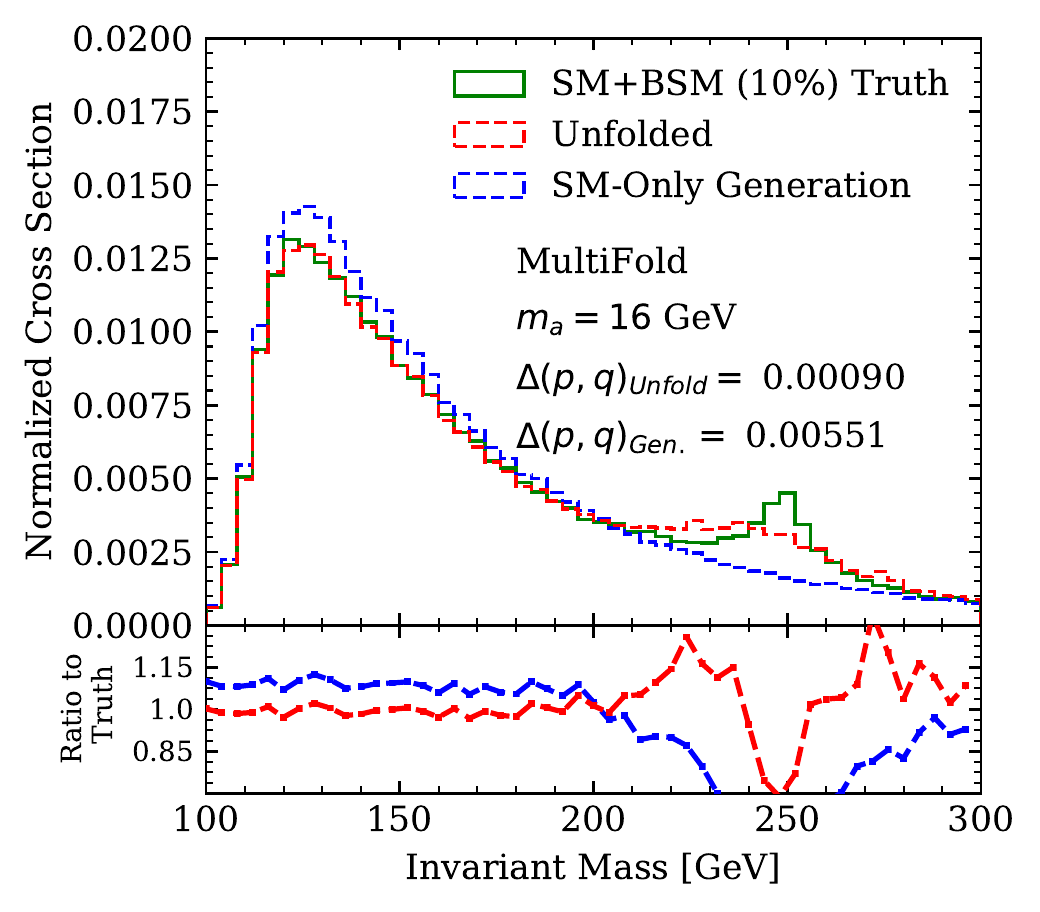}
  \includegraphics[height=3.5cm]{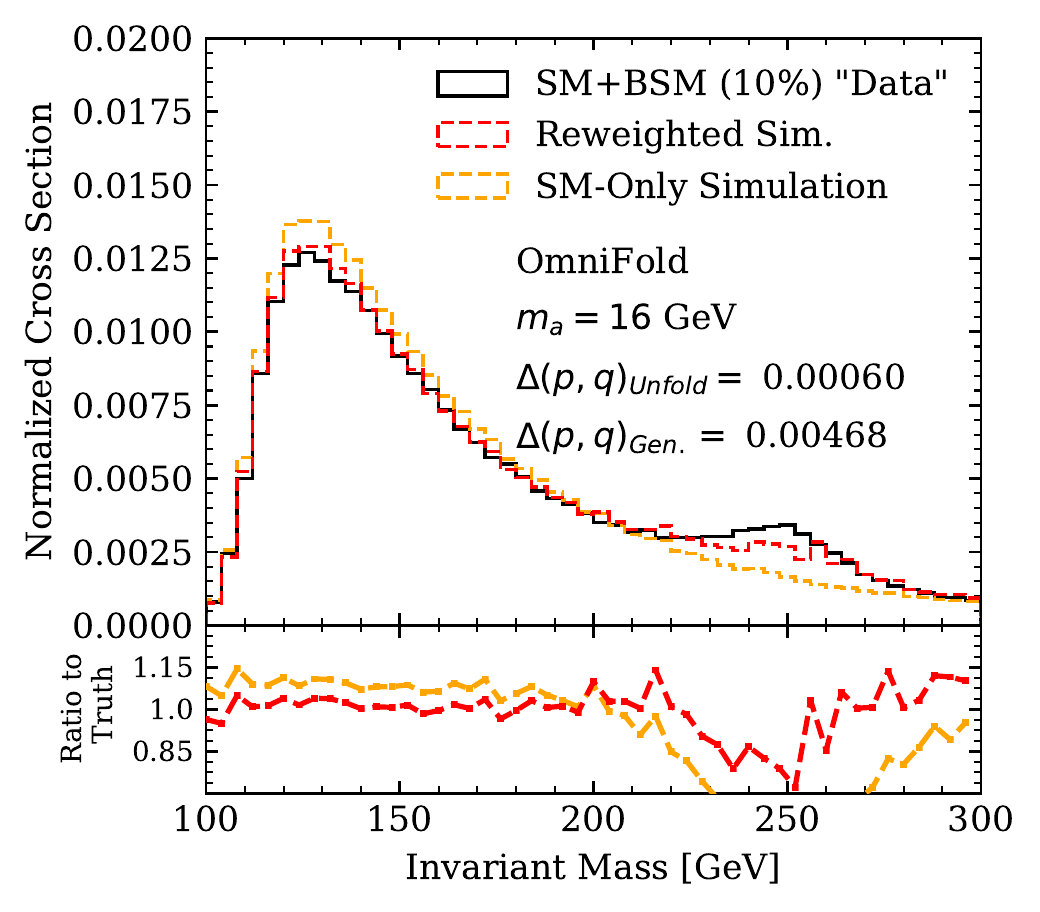}
  \includegraphics[height=3.5cm]{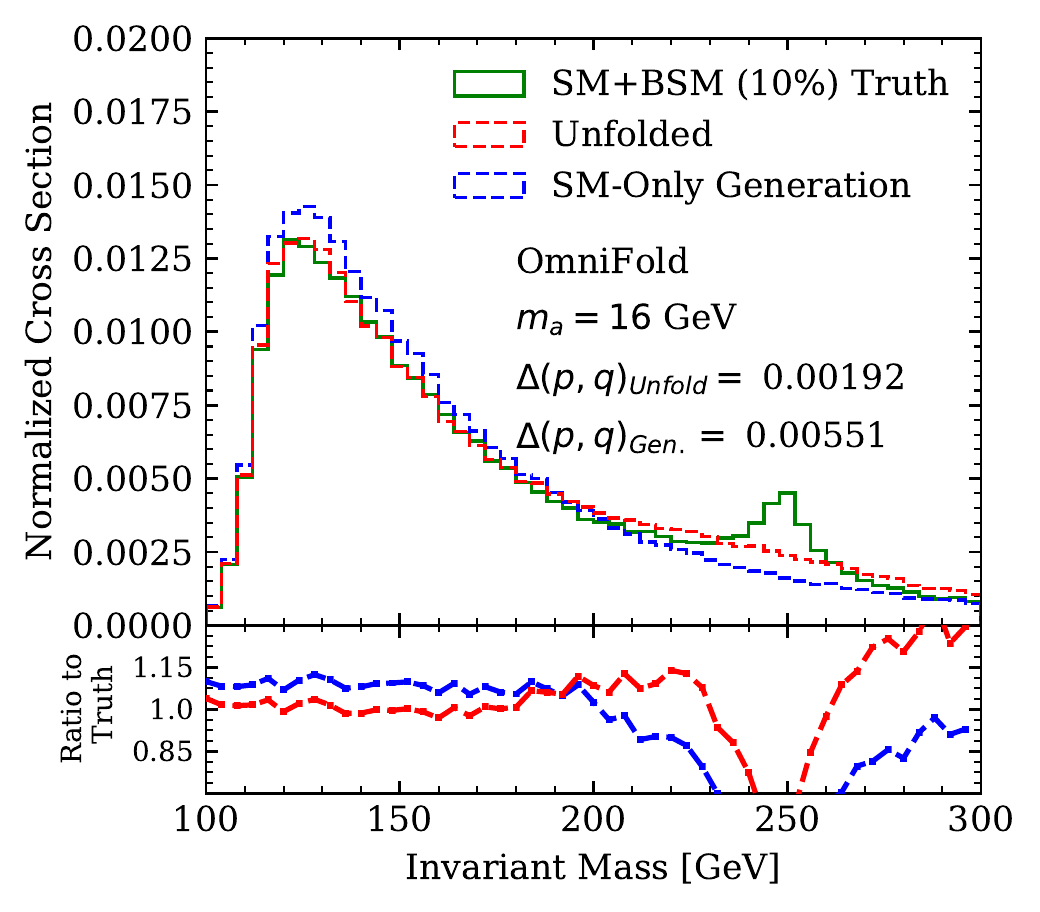}
  \caption{Distributions of the $Z$+jet invariant mass spectrum for both \multifold~(top row)~and \omnifold~(bottom row).  Distributions are shown for both detector-level (Data and Simulation) and truth-level (Truth and Generation) values.  The Truth and Data distributions are a combination of 180,000 \textsc{Pythia}~8 $Z$+jet events and 20,000 $h\rightarrow Za, a\rightarrow gg$, where $m_h$ = 250 GeV and $m_a$ = 16 GeV.  The Generation and Simulation are 200,000 SM-only events.  The weights are taken after 5 iterations of the respective unfolding procedure.  The triangular discriminator~\cite{Harrison:1998yr,Hocker:2007ht,Triangular_discrim} $\Delta (p,q) = \int d\lambda \frac{(p(\lambda) - q(\lambda))^2}{p(\lambda) + q(\lambda)}$ is used to quantify the difference between distributions.}
  \label{fig:Higgs250}
\end{figure}

Figure~\ref{fig:Higgs250} shows the detector-level and truth-level distributions of $Z$+jet invariant mass before and after unfolding.  At detector-level, which corresponds to the first step in an iteration of the \omnifold~method, the distributions exhibit good agreement after unfolding: the height and width of the mass peak are reproduced accurately, especially in the \multifold~case.  At truth-level, the peaks are not reproduced as sharply.  In the \multifold~case, the height and width of the peak are similar to that seen at detector level, and in the \omnifold~case, the peak is considerably broader.

\begin{figure}
  \centering
  \includegraphics[width=0.45\textwidth]{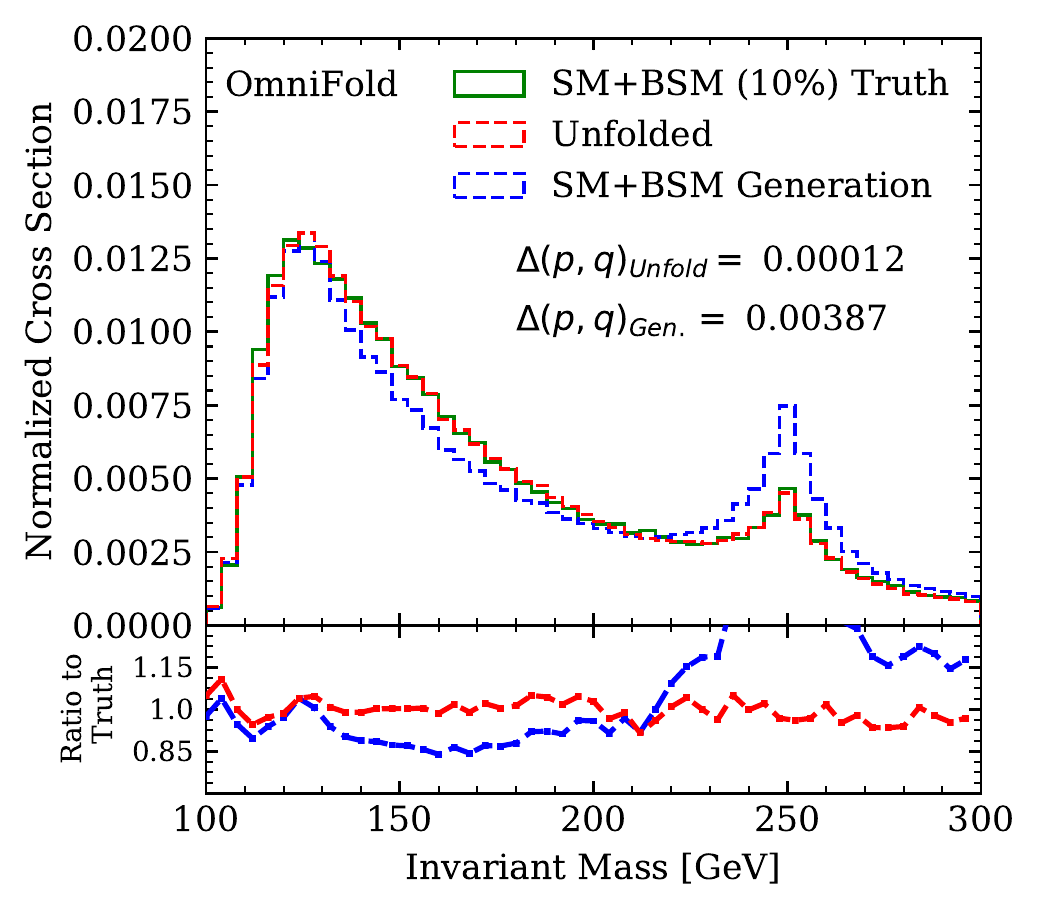}
  \caption{Truth-level distribution of $Z$+jet invariant mass for the case that \omnifold~is performed with BSM events in Generation.  The BSM event included in the Generation were drawn from events with $m_h$ = 125 GeV and $m_h$ = 250 GeV.  The weights are taken after 5 \omnifold~iterations.}
  \label{fig:Higgs250_BSMGen}
\end{figure}

Part of the broadening is an inherent challenge with non-trivial resolutions and limited statistics.  The truth-level peak quality can be recovered by modifying the Generation.  We are free to choose whatever Generation we want as OmniFold is a maximum likelihood estimator that is prior-independent.  However, the closer the prior is to the data, the more accurate the unfolding will be with finite statistics.  To test this idea, the same Truth sample was used as above, with 180,000 SM events and 20,000 $h\rightarrow Za, a\rightarrow gg$, where $m_h$ = 250 GeV and $m_a$ = 16 GeV.  However, now the Generation was taken to include 200,000 SM events, 10,000 $h\rightarrow Za, a\rightarrow gg$ events with $m_h$ = 125 GeV for each of $m_a$ = 0.5, 1, 2, 4, 8, and 16 GeV, and 10,000 $h\rightarrow Za, a\rightarrow gg$ events with $m_h$ = 250 GeV for each of the same $m_a$ values, for a total of 320,000 events.  The truth-level results of unfolding with the same \omnifold~setup discussed above are shown in Fig.~\ref{fig:Higgs250_BSMGen}.  Here, both the height and weight of the truth-level peak are reproduced well by the reweighted sample. The fact that this works well, when an application of OmniFold with SM-only events did not, shows the importance of sufficiently covering the relevant regions of phase space.

\begin{figure}
  \centering
  \includegraphics[width=0.45\textwidth]{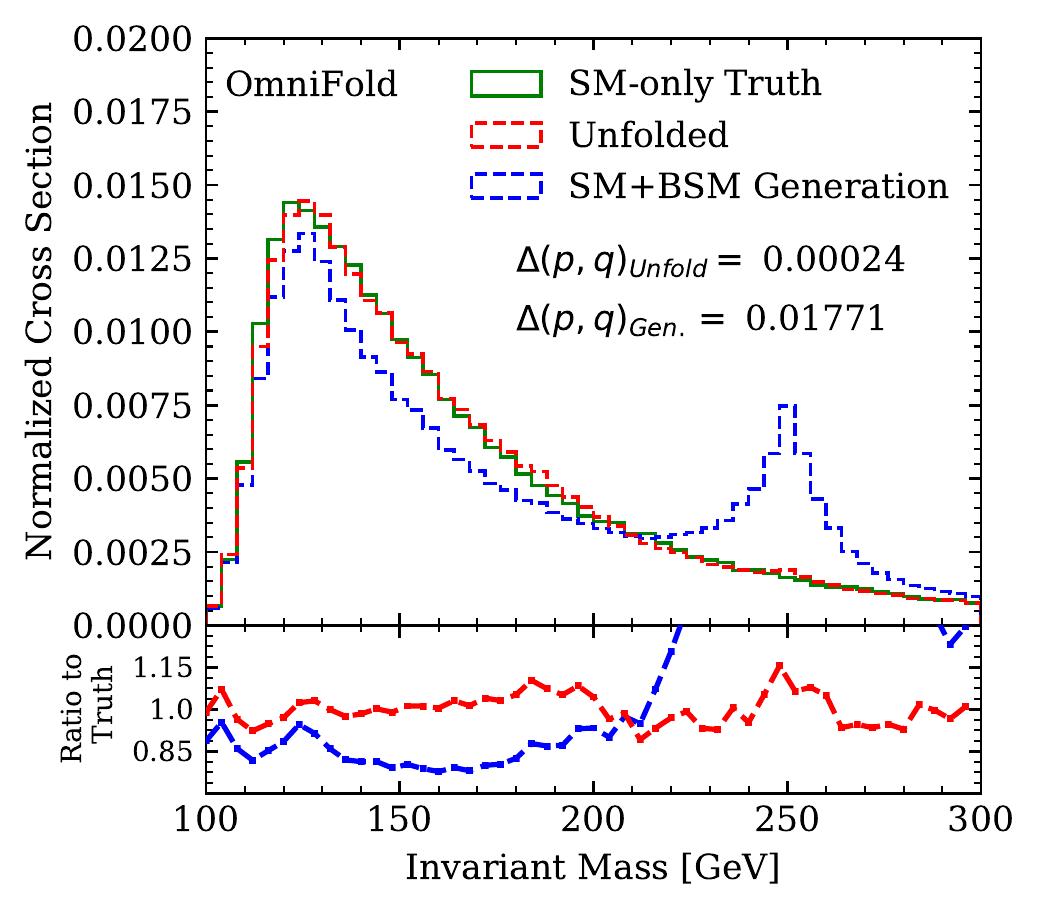}
  \caption{Truth-level distribution of $Z$+jet invariant mass for the case that \omnifold~is performed with BSM events in Generation despite a lack of BSM events in Truth.  The BSM event included in the Generation were drawn from events with $m_h$ = 125 GeV and $m_h$ = 250 GeV.  The weights are taken after 5 \omnifold~iterations.}
  \label{fig:Higgs250_BSMGen_0Perc}
\end{figure}

Adding BSM physics to the Generation sample begs the question of what the invariant mass distribution would look like after unfolding if the Data \textit{does not} itself contain BSM physics.  To test this, the same Generation sample with 320,000 events from the preceding paragraph is used again, but Data and Truth are taken to be 200,000 SM-only events.  The \omnifold~method is applied in the same way as above, and the resulting $Z$+jet invariant mass distributions are shown in Fig.~\ref{fig:Higgs250_BSMGen_0Perc}.  The bump in the unfolded distribution has been eliminated despite the fact that almost 40\% of events in the Generation sample were drawn from BSM samples.  To optimize the unfolding, care must be taken when choosing the BSM models to include in the Generation sample as well as when choosing the number of BSM events to include -- manipulating these parameters effectively corresponds to choosing different priors for what is expected in the Data.

This section has demonstrated that \omnifold~can qualitatively preserve a relatively large\footnote{In fact, the amount of signal is so large, that it would result in a significant detection from the cross-section alone, which is well-known for $Z$+jets.  We revisit this in Sec.~\ref{sec:conclusions}.} and prominent resonant signature from the data.  A sideband technique could then be used to perform a search with these data.  In the next section, we will explore the ability of \omnifold~to precisely preserve the phase space so that a multivariate classifier could be used for a search with the unfolded data.

\section{Exotic Higgs Decay}
\label{sec:results}

The signal in Sec.~\ref{sec:Higgs250} was sufficiently prominent that it could be effectively searched for with a bump hunt in the $Z$+jet invariant mass.  Not all new physics processes can be searched for with such a simple approach.  
To explore such a scenario, we consider $m_h = 125$ GeV.  In this case, the signal bump is near the background peak and so additional features beyond just the $Z$+jet invariant mass are required.  We explore the possibility of performing a \textit{model-dependent} search that uses dedicated BSM vs. SM discriminating variables.  If \omnifold~effectively unfolds the full phase space, it should be possible to use any combination of variables in unfolded data.  


\subsection{
Unfolding with \multifold}
\label{multifold_125}

First, we can consider the case of \multifold~with two working points for BSM physics:
\begin{itemize}
    \item 0.1\% of Data and Truth events are BSM physics, with $m_a = 16$ GeV.
    \item 10\% of Data and Truth events are BSM physics, with $m_a = 16$ GeV.
\end{itemize}
For each working point, the Data, Truth, Simulation, and Generation sets will be once again be 200,000 events.  In each case, the Simulation and Generation sets are drawn from the SM-only \textsc{Pythia}~8 sample.  The SM events in Data and Truth are also drawn from the \textsc{Pythia}~8 sample, but no SM event can be used in both Data and Simulation.

\begin{figure}
  \centering
  \includegraphics[height=3.5cm]{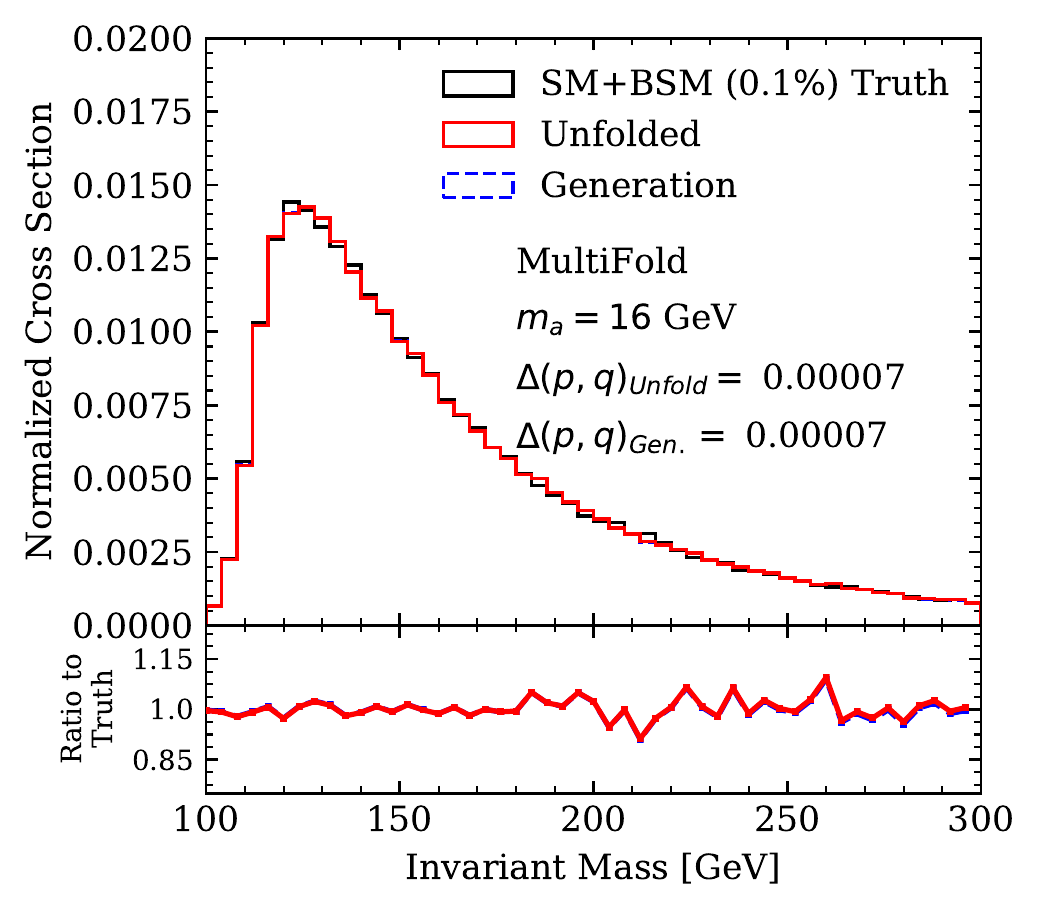}
  \includegraphics[height=3.5cm]{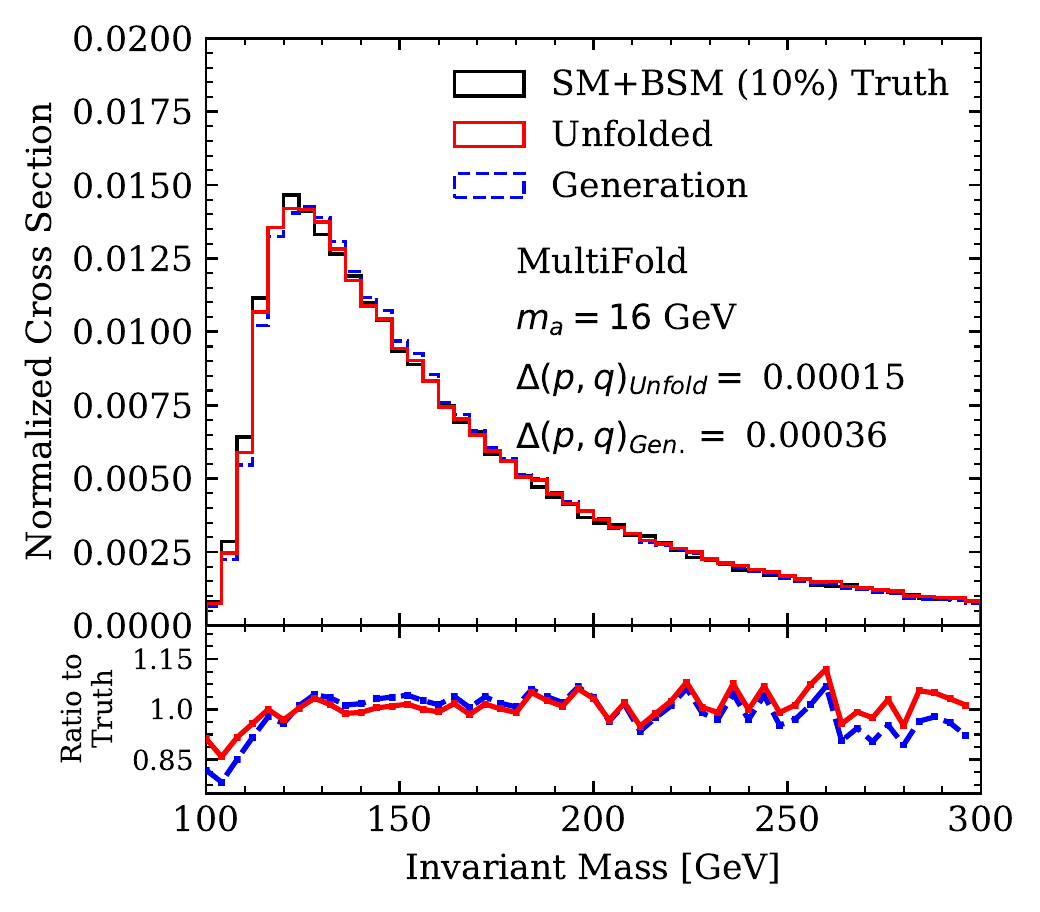}
  \includegraphics[height=3.5cm]{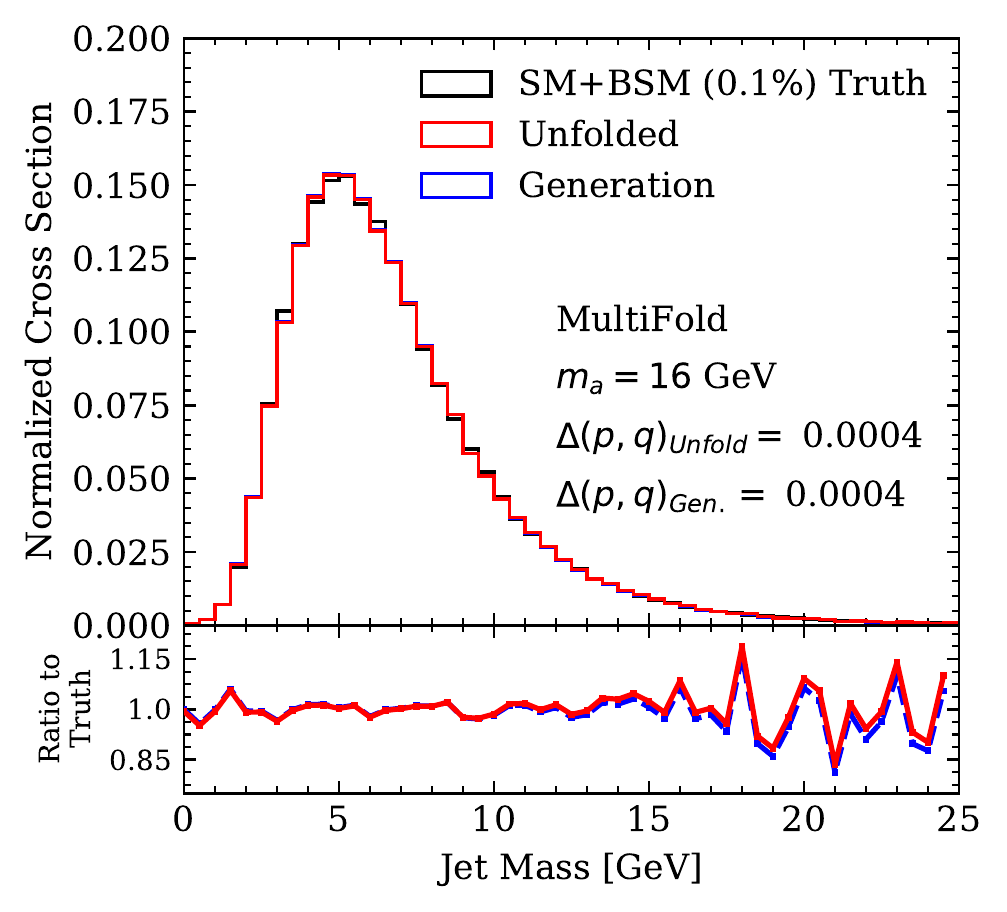}
  \includegraphics[height=3.5cm]{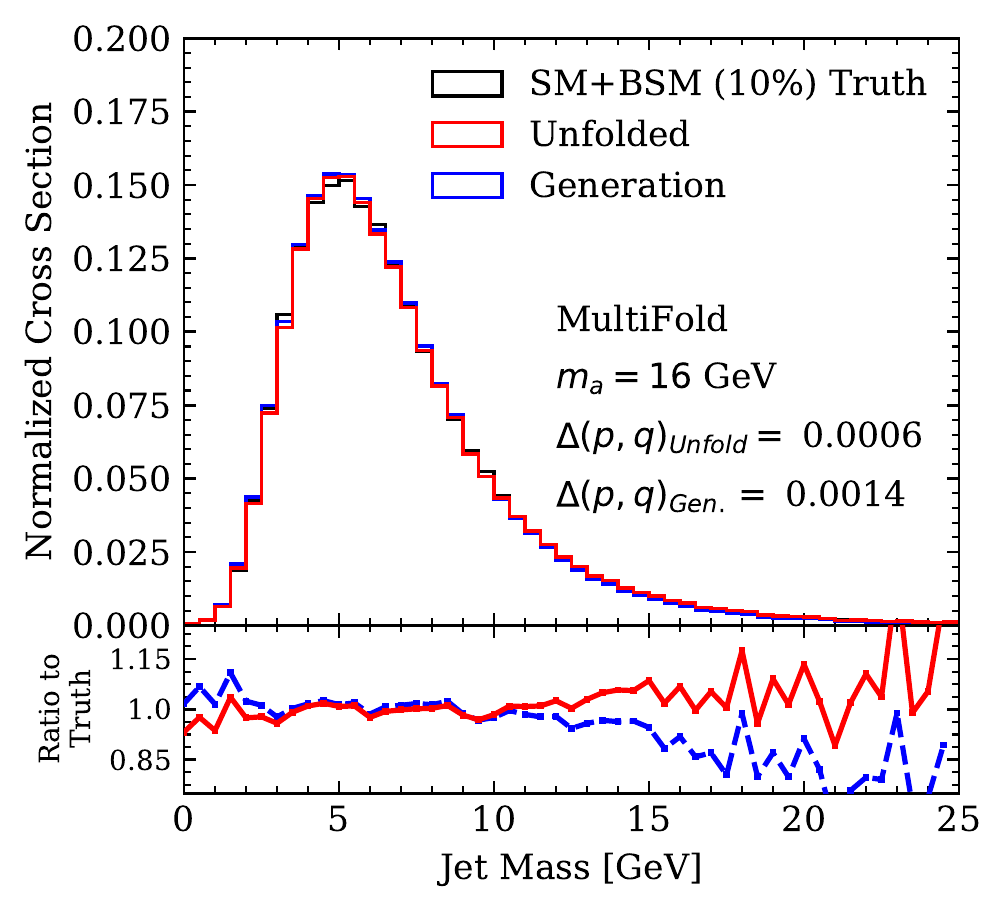}
  \includegraphics[height=3.5cm]{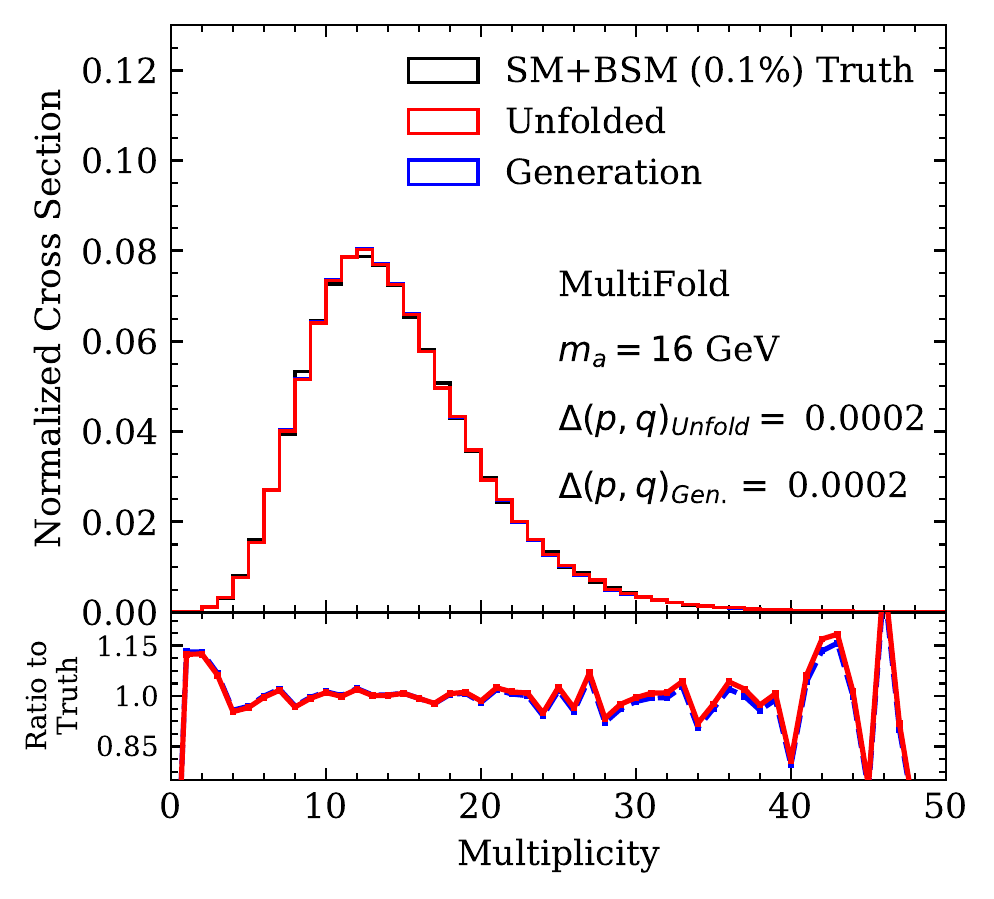}
  \includegraphics[height=3.5cm]{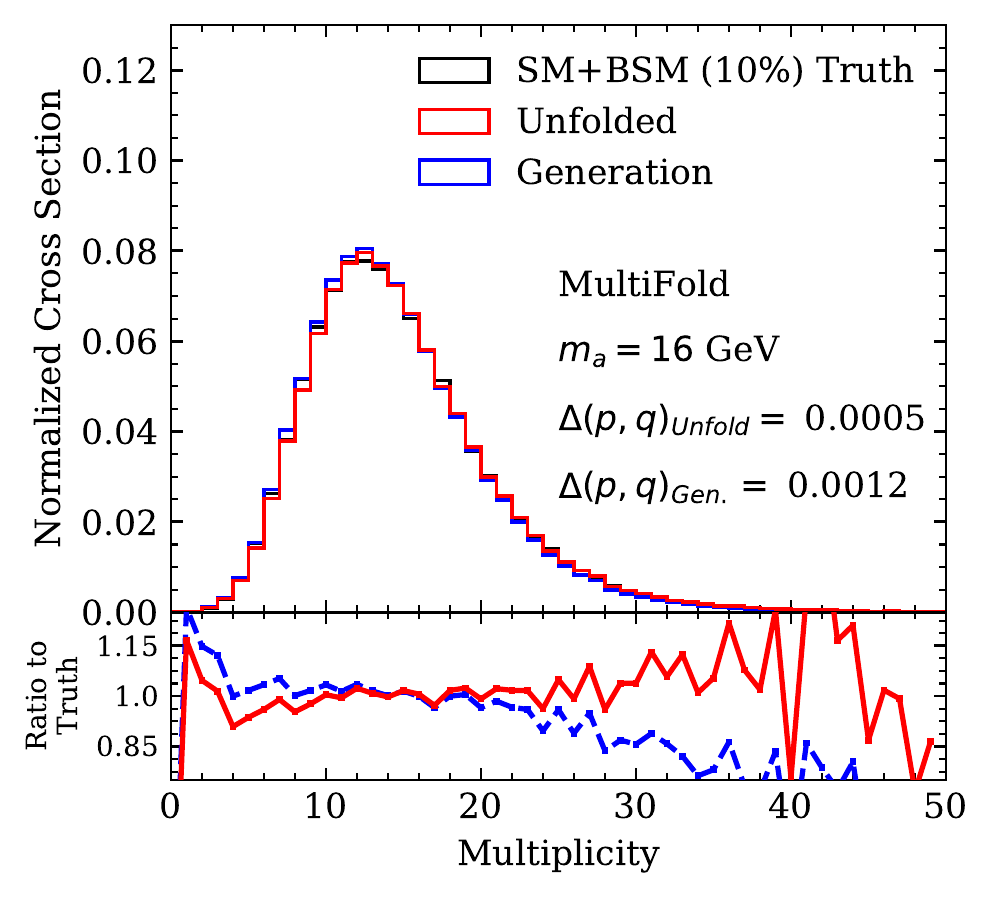}
  \caption{Truth, Generation, and unfolded Generation distributions for the \multifold~case, where BSM $h\rightarrow Za, a\rightarrow gg$ events have been included in the Truth, but not the Generation.  Standard Model events in these samples come from \textsc{Pythia}~8 $Z$+jets simulation.  In the left column, 200 out of 200,000 Truth events come from the BSM sample, and in the right column, 20,000 out of 200,000 Truth events are BSM physics.  Distributions are given for the invariant mass of the $Z$+jet, the jet mass, and the jet multiplicity.  The ratios of the Generation distributions are given to Truth for each plot.  The weights are taken after 5 iterations of \multifold.}
  \label{fig:multi_16GeV}
\end{figure}

The distributions of $Z$+jet invariant mass, jet mass, and jet multiplicity for Truth, Generation, and unfolded Generation are shown in Fig.~\ref{fig:multi_16GeV}.  The impact of a 0.1\% signal is difficult to detect in these one-dimensional histograms and \multifold~has correspondingly left the phase space mostly untouched.  For the 10\% signal, \multifold~clearly improves the agreement of the distributions.  Similar trends hold for alternative $m_a$ values as well (not shown).

The ratio panels in Fig.~\ref{fig:multi_16GeV} show that \multifold~is qualitatively able to encode BSM physics in the unfolded data.  To probe this in greater detail, we emulate a model-dependent search by training a fully supervised classifier to distinguish $Z$+jets events from the $m_a=16$ GeV signal.  A sample of 90,000 SM and 90,000 BSM events was used for training, with 30\% randomly held out as a validation set.  The neural network has the same inputs and architecture as the one used for \multifold\footnote{It is important to note that in the \multifold~case, the neural network is trained to distinguish between Data and Simulation, whereas the discriminator neural network is trained to distinguish truth-level SM events from BSM events.}.  If \multifold~preserves the complete phase space, then any threshold cut on this classifier should have the same efficiency with the unfolded data as it does with the Truth.

\begin{figure}
  \centering
  \includegraphics[height=5.5cm]{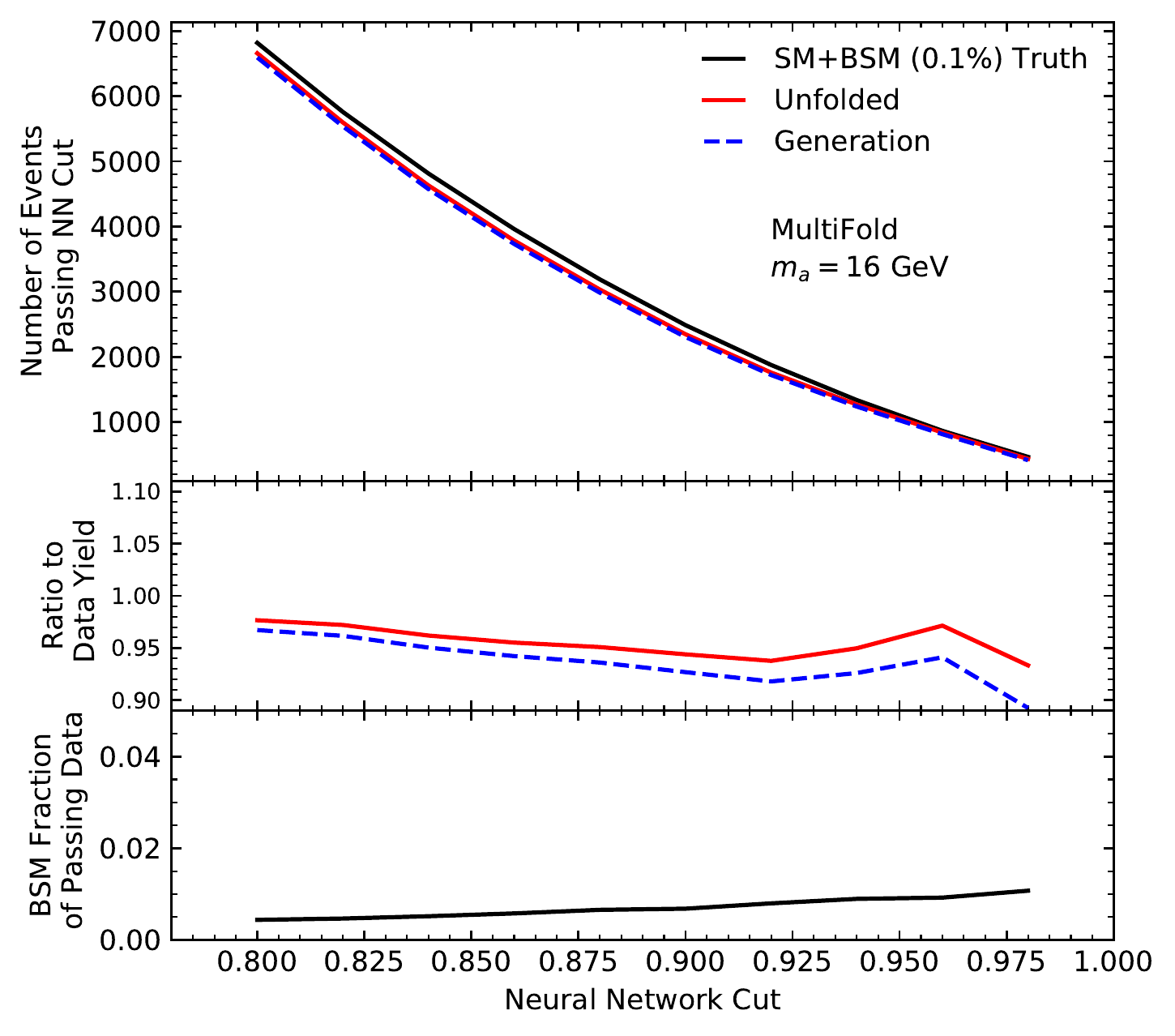}
  \includegraphics[height=5.5cm]{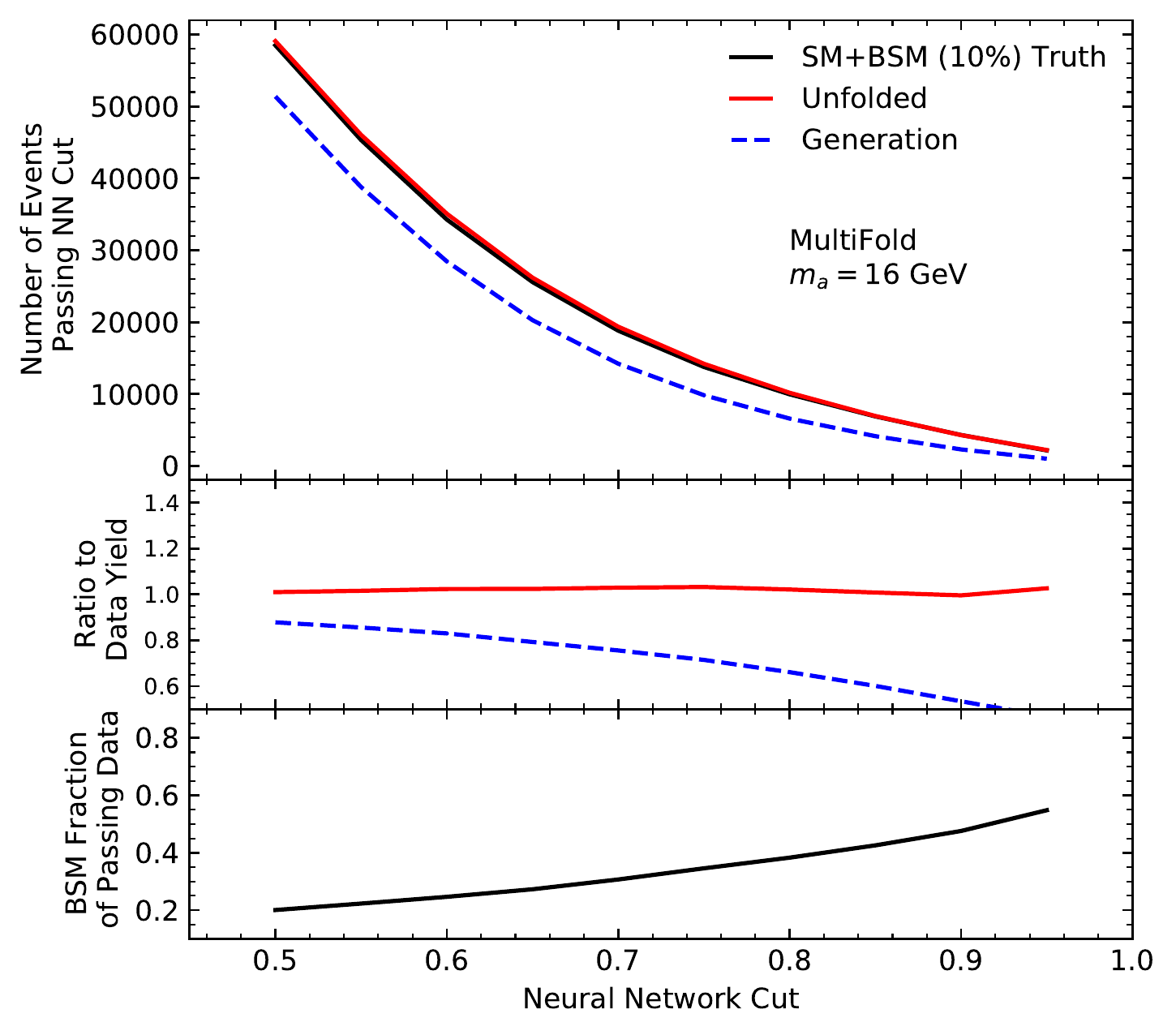}
  \caption{The number of Truth, Generation, and unfolded Generation events passing a cut on the neural network score, as a function of the cut value, in the case that 0.1\% of the data comes from BSM physics (top) and 10\% of the data comes from BSM physics (bottom).  The unfolding was performed with \multifold.  The neural network was specifically trained to distinguish SM from BSM physics.  The middle segment of both plots shows the ratio of the pre- and post-unfolding Generation yields to the Truth yield.  The yield from the pre-unfolding Generation is not expected to agree well with the Truth, as there are no BSM events.  However, if the most \textit{BSM-like} events get upweighted by the unfolding, then the weighted sum of passing events increases.  In a fully accurate unfolding the unfolded yield would match the yield from Truth.  The bottom segment shows the fraction of Truth events that are BSM events at truth-level; as the cut value approaches 1, the events passing the selection must be more BSM-like, which is why the BSM-purity increases as a function of cut value.}
  \label{fig:multifold_discriminator}
\end{figure}

The number of Truth, Generation, and unfolded Generation events passing a cut on the neural network score, as a function of the cut value, is shown in Fig.~\ref{fig:multifold_discriminator}.  In both the 0.1\% and 10\% signal case, the number of events in the reweighted samples more closely matches the Truth than the raw Generation samples.  The agreement between the reweighted sample and Truth in the 10\% case is an impressive achievement, as the Truth and Unfolded yields after the application of the NN score cut is stable at one.  The NN score is a specialized value that was not used in training, so it is clear that in this case, the most BSM-like events are being up-weighted to an appropriate degree.  In contrast, the unfolding has not up-weighted the BSM events enough in the 0.1\% case, highlighting the difficulty of working with such a small signal.

\subsection{
Unfolding with \omnifold}
\label{sec:omnifold_125}

An investigation similar to the \multifold~case can be performed with \omnifold.  The same $m_a$ and contamination values are investigated.

\begin{figure}
  \centering
  \includegraphics[height=3.5cm]{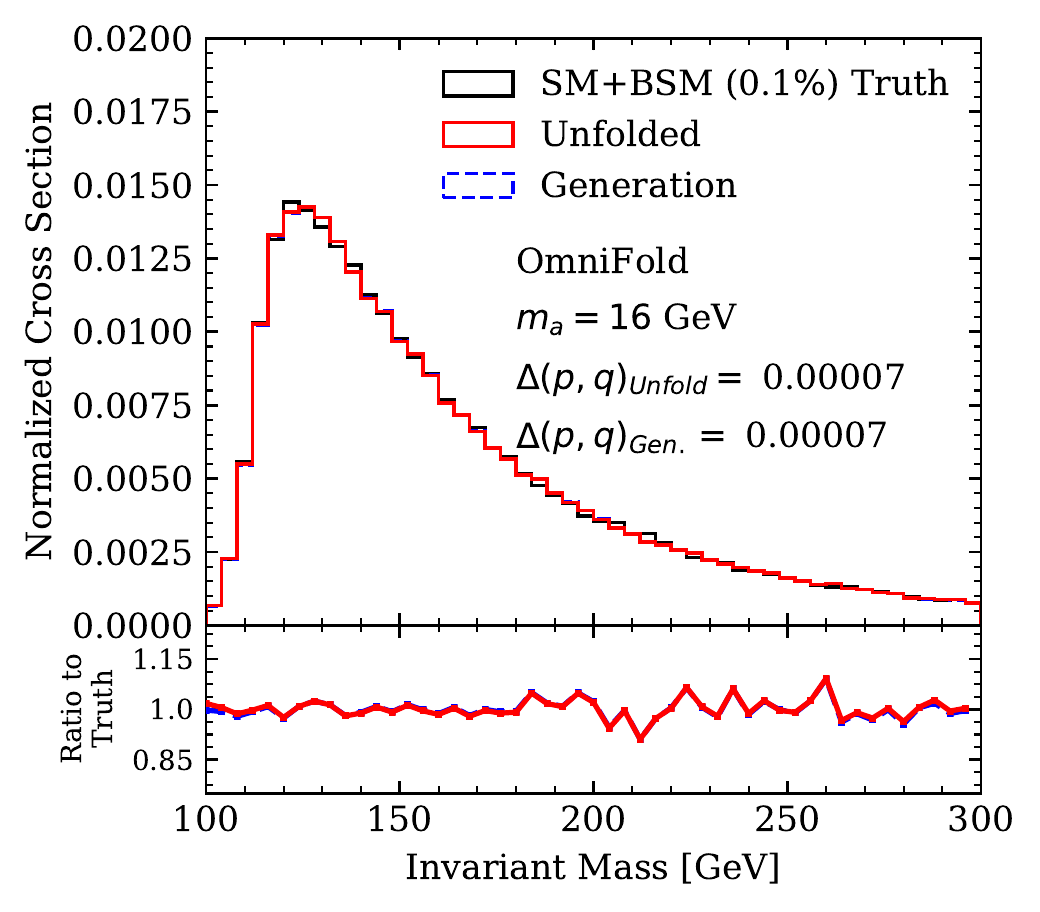}
  \includegraphics[height=3.5cm]{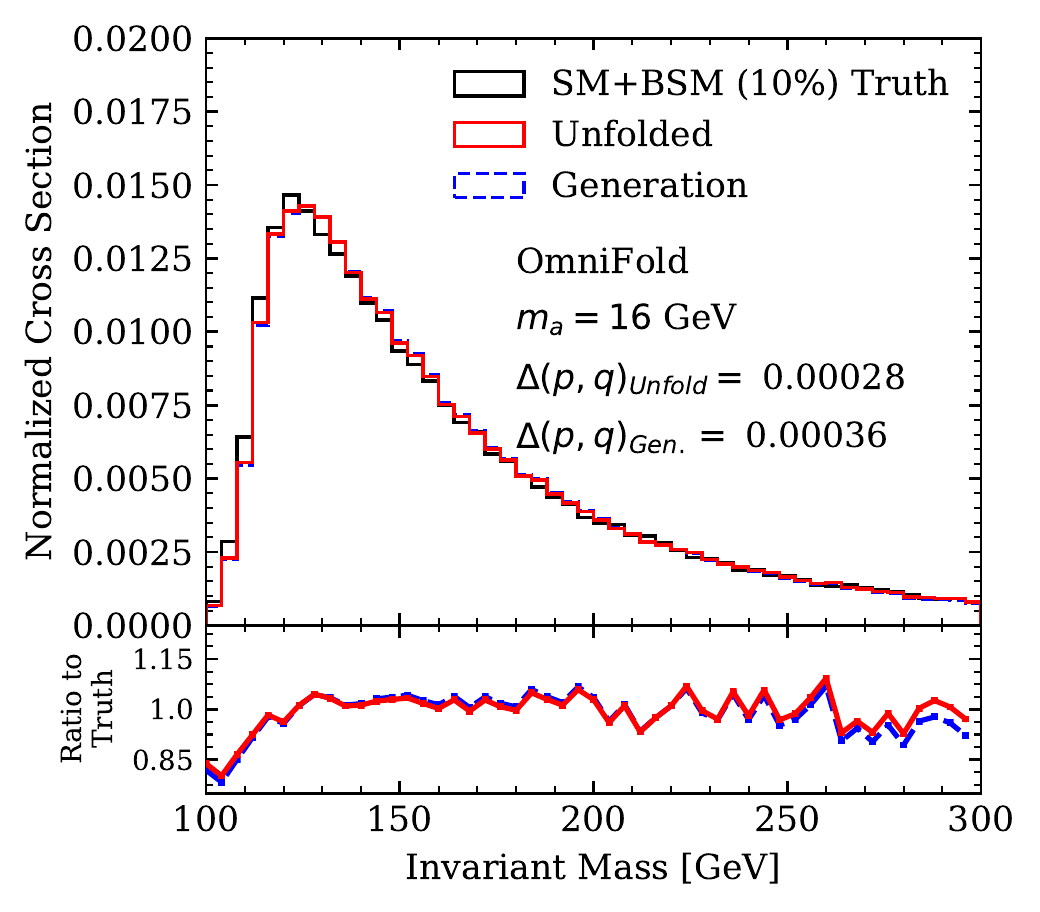}
  \includegraphics[height=3.5cm]{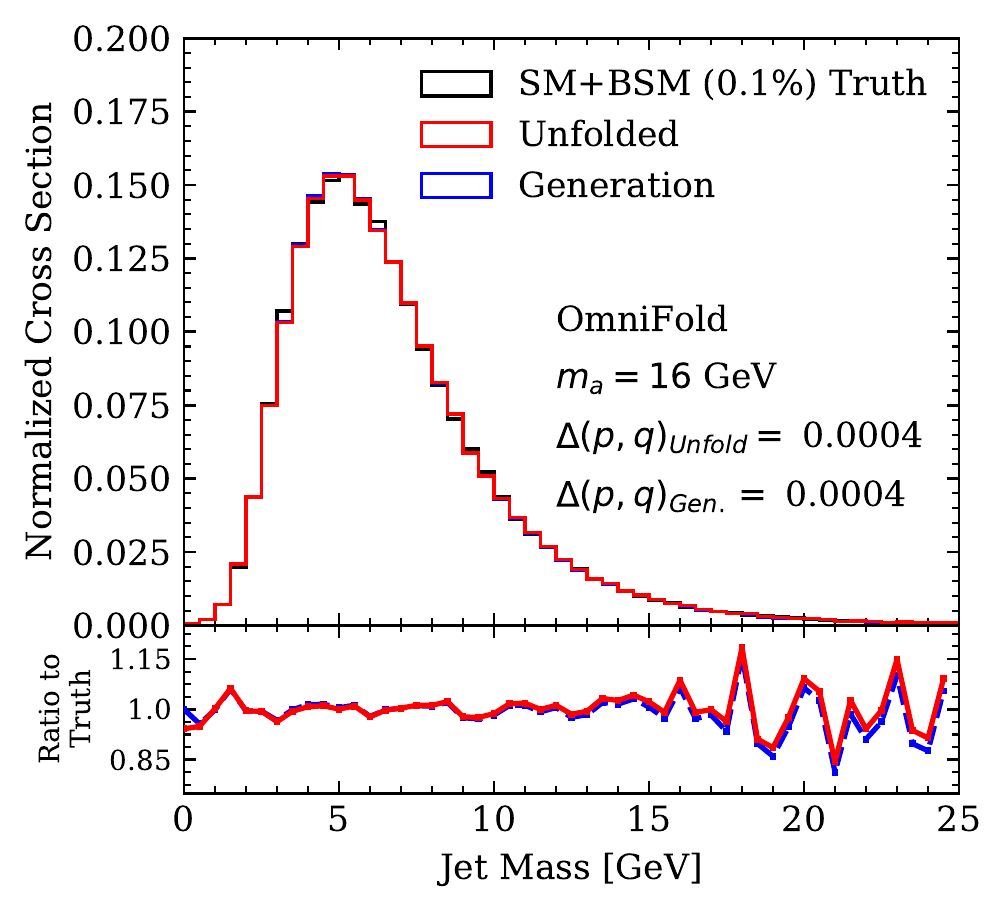}
  \includegraphics[height=3.5cm]{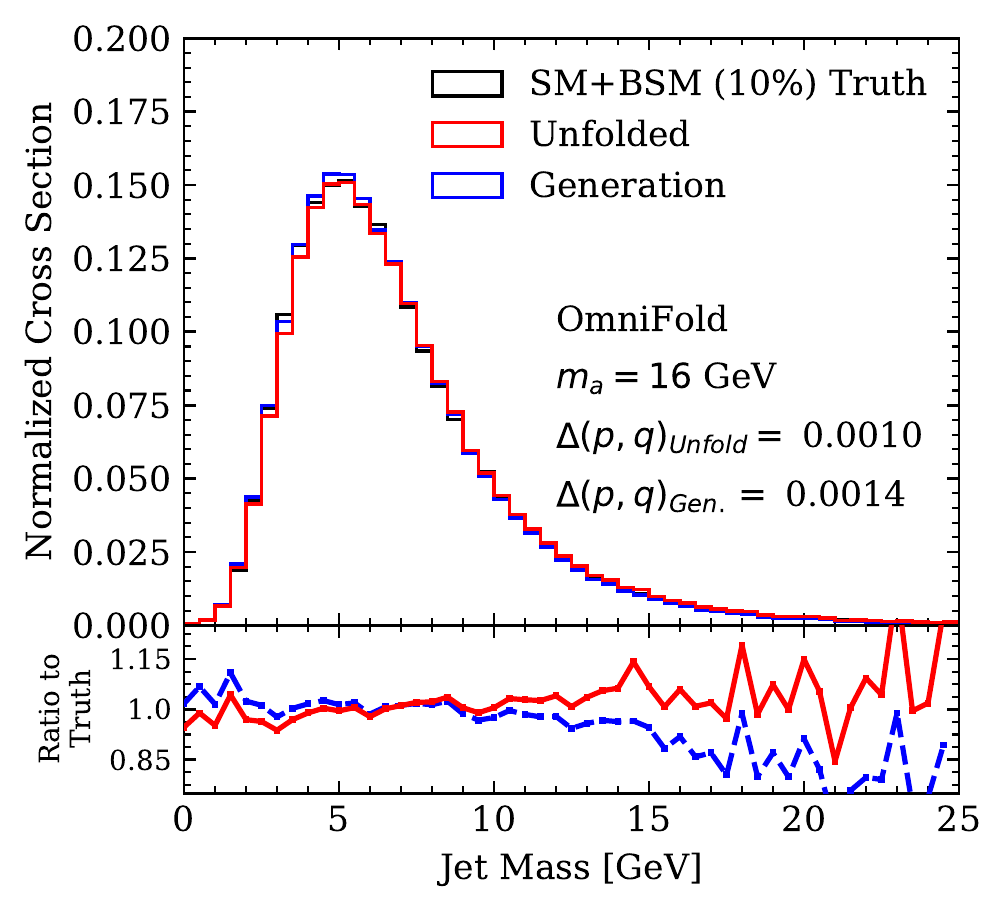}
  \includegraphics[height=3.5cm]{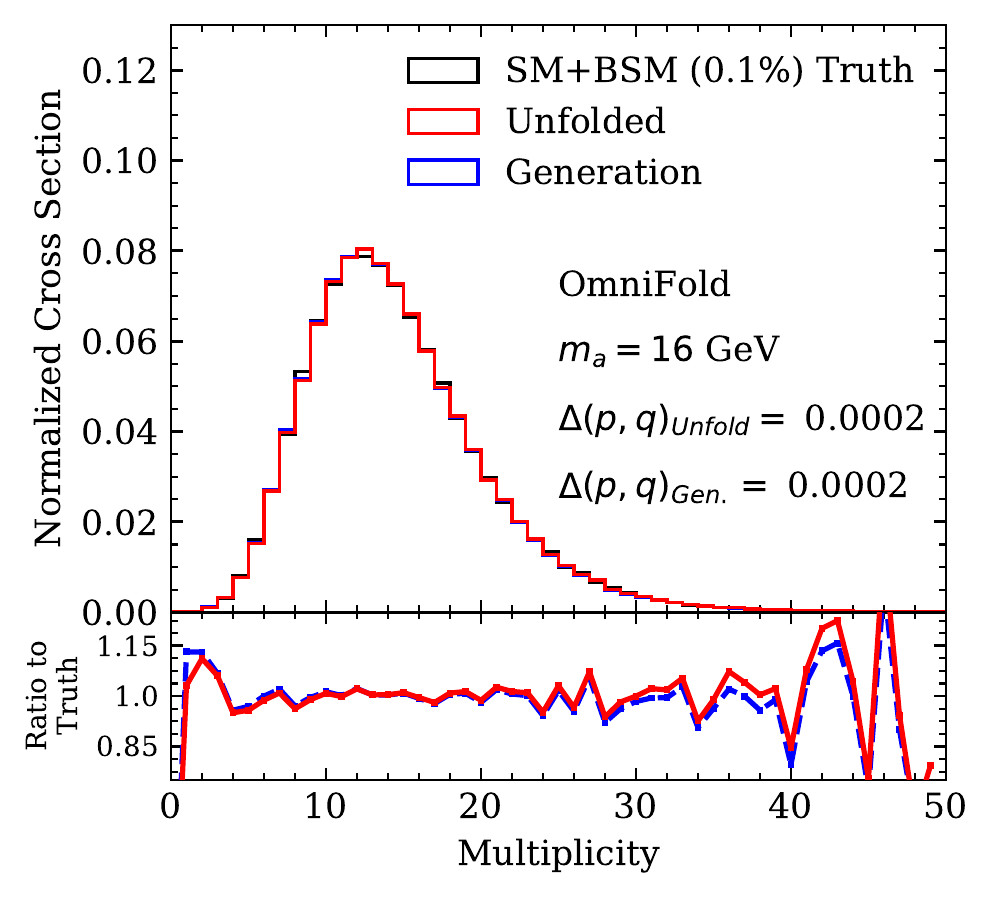}
  \includegraphics[height=3.5cm]{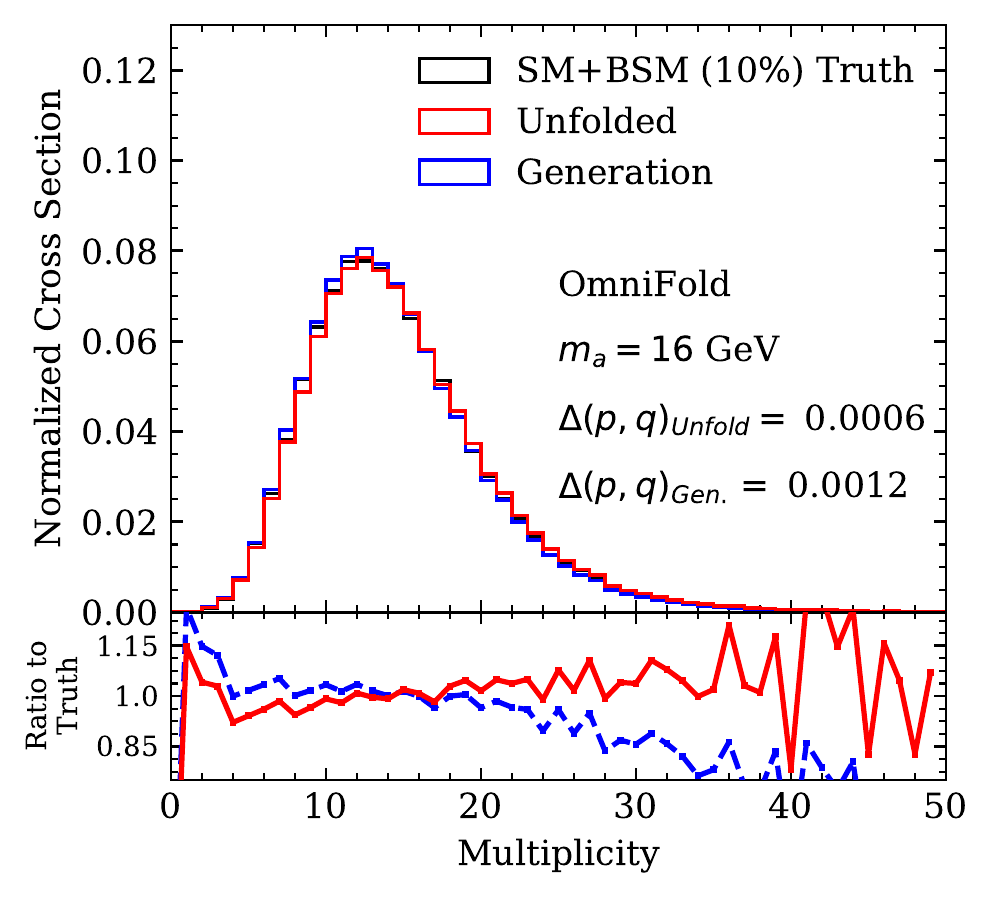}
  \caption{Truth, Generation, and unfolded Generation distributions for the \omnifold~case, where BSM $h\rightarrow Za, a\rightarrow gg$ events have been included in the Truth, but not the Generation.  Standard Model events in these samples come from \textsc{Pythia}~8 $Z$+jets simulation.  In the left column, 200 out of 200,000 Truth events come from the BSM sample, and in the right column, 20,000 out of 200,000 Truth events are BSM physics.  Distributions are given for the invariant mass of the $Z$+jet, the jet mass, and the jet multiplicity.  The ratios of the Generation distributions are given to Truth for each plot.  The weights are taken after 3 iterations of \omnifold.}
  \label{fig:omni_16GeV}
\end{figure}

The distributions of $Z$+jet invariant mass, jet mass, and jet multiplicity for Truth, Generation, and unfolded Generation are shown in Fig.~\ref{fig:omni_16GeV}.  The performance in these plots is similar to but slightly worse than that observed in Fig.~\ref{fig:multi_16GeV}.

\begin{figure}
  \centering
  \includegraphics[height=5.5cm]{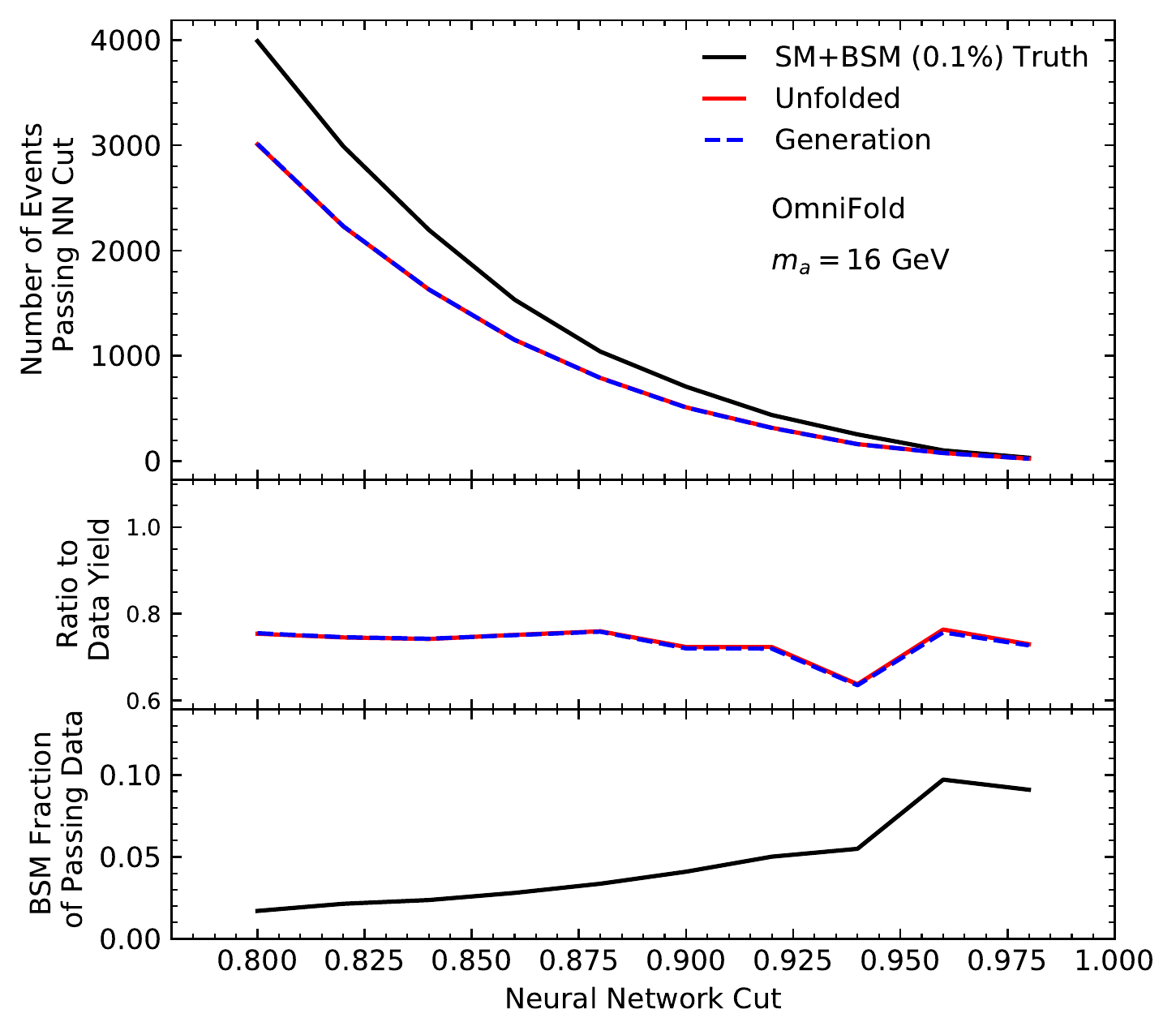}
  \includegraphics[height=5.5cm]{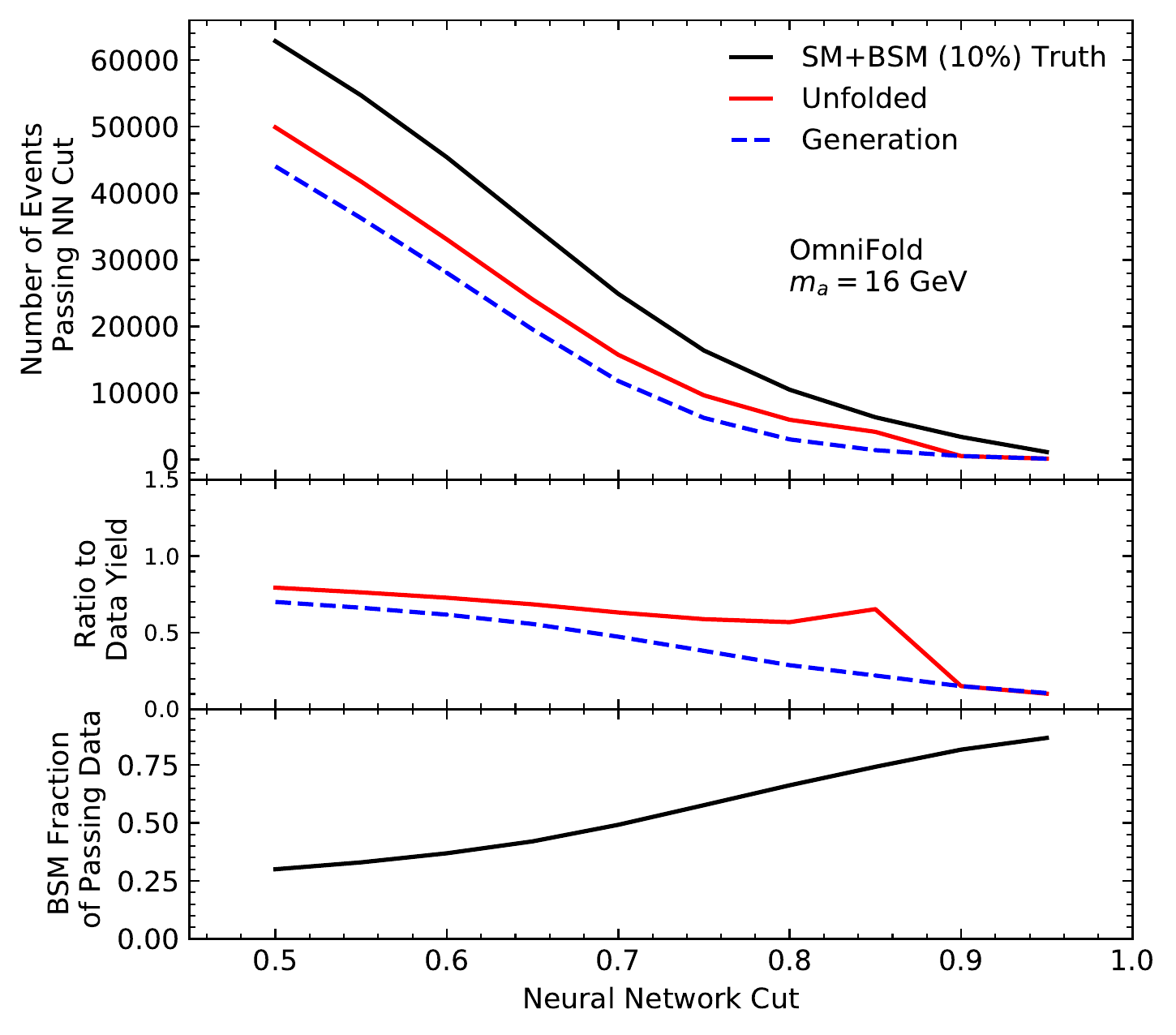}
  \caption{The number of Truth, Generation, and unfolded Generation events passing a cut on the PFN score, as a function of the cut value, in the case that 0.1\% of the data comes from BSM physics (top) and 10\% of the data comes from BSM physics (bottom).  The unfolding was performed with \omnifold.  The PFN was specifically trained to distinguish SM from BSM physics.  The middle segment of both plots shows the ratio of the pre- and post-unfolding Generation yields to the Truth yield.  The bottom segment shows the fraction of Truth events that are BSM events at truth-level.}
  \label{fig:omnifold_discriminator}
\end{figure}

It is also possible to train a PFN to distinguish SM from BSM events.  This PFN is set up in the same way as the PFN used for \omnifold, but it is trained to discriminate a sample of 90,000 SM from 90,000 BSM events.  Here, the area under the ROC curve is 0.94, achieving superior discrimination to the neural network described in Sec.~\ref{multifold_125} (AUC of 0.73).  Figure~\ref{fig:omnifold_discriminator} shows the number of Truth, Generation, and unfolded Generation events that pass cuts on the PFN score.  If this figure is compared to Fig.~\ref{fig:multifold_discriminator}, it is evident that the post-cut yields in the \omnifold~case do not agree with truth as well as in the \multifold~case.  This is due to the challenges discussed in Sec.~\ref{sec:limitations}.  When specifically trained to discriminate SM from BSM events, the PFN is highly accurate.  The Generation sample poorly populates the very BSM-like region of this discriminator.  Especially in the case of 0.1\% contamination, the unfolded dataset would not enable a discovery of new physics, as the weights do not strongly affect the post-cut yields.

\subsection{
Including BSM Physics in Generation}
\label{sec:BSMGen}

\begin{figure}
  \centering
  \includegraphics[height=3.5cm]{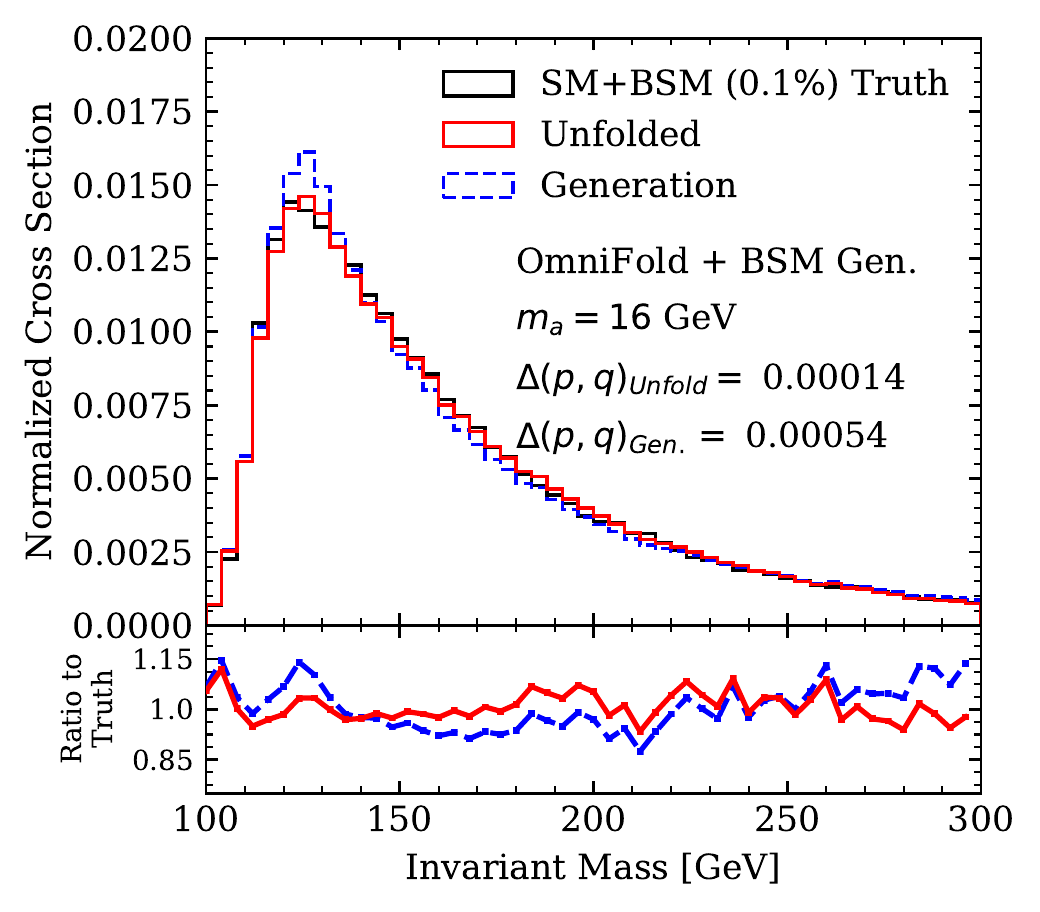}
  \includegraphics[height=3.5cm]{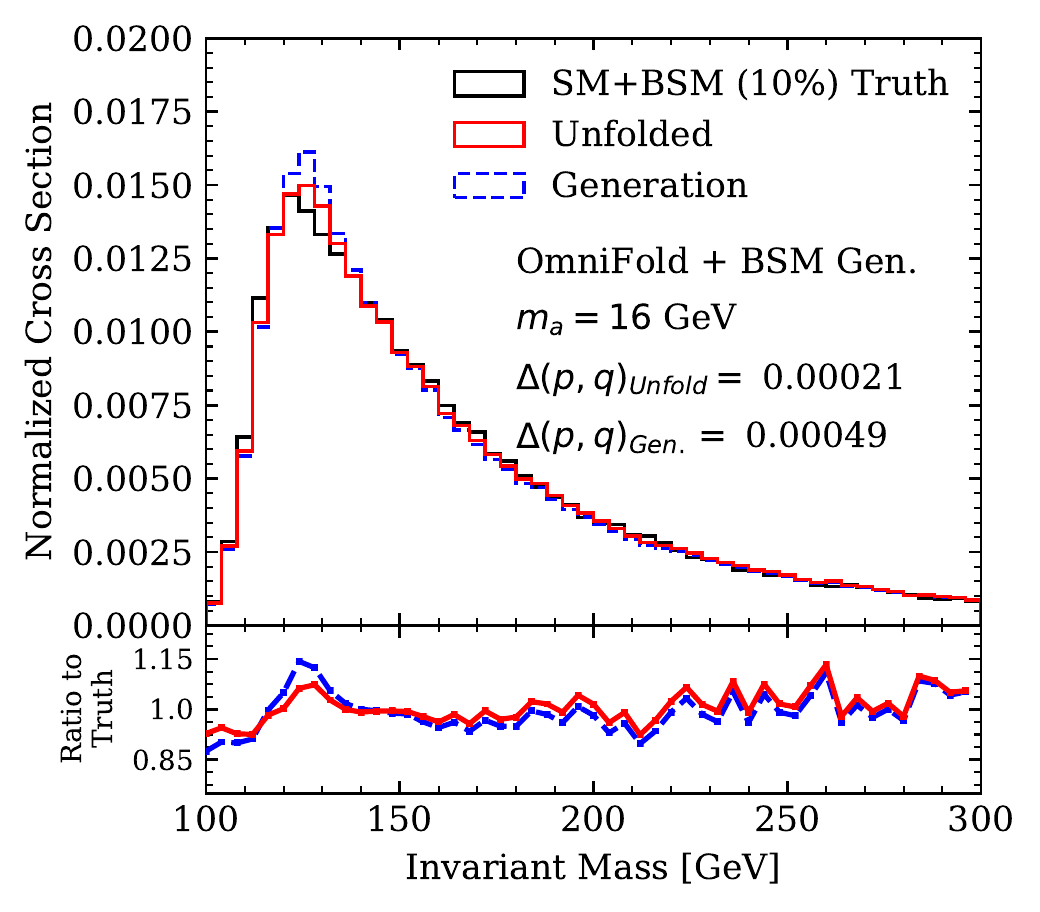}
  \includegraphics[height=3.5cm]{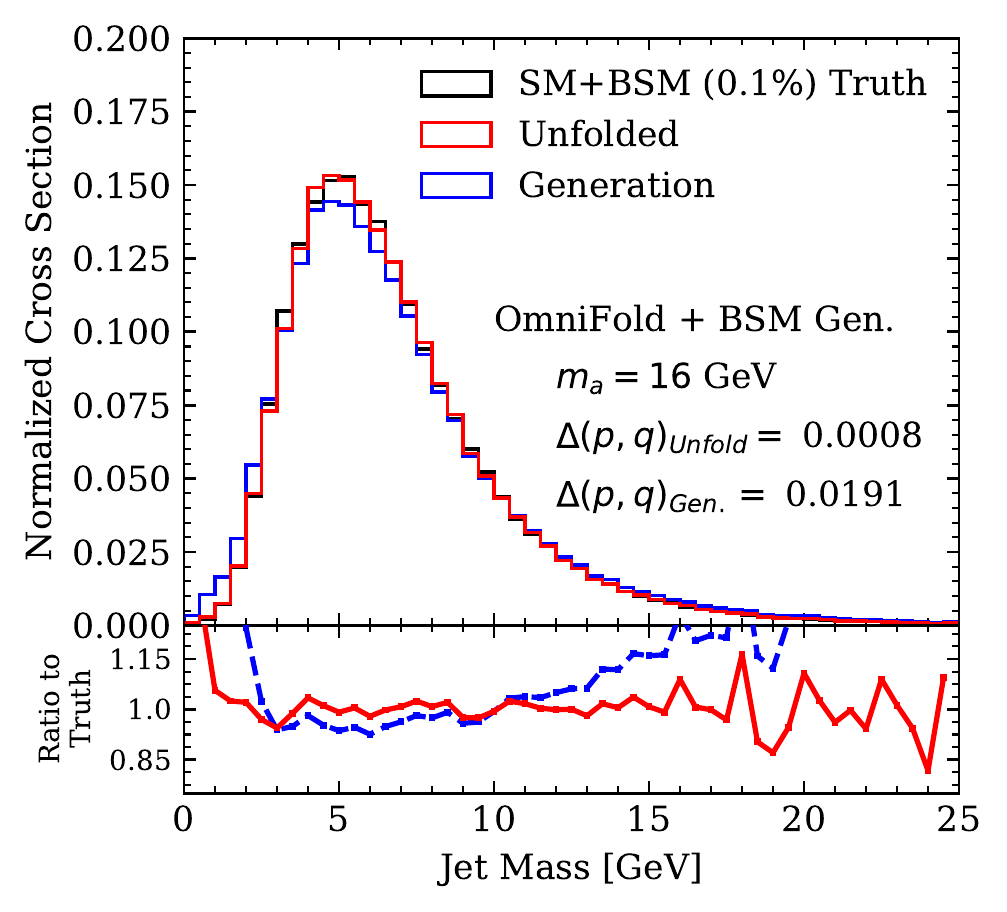}
  \includegraphics[height=3.5cm]{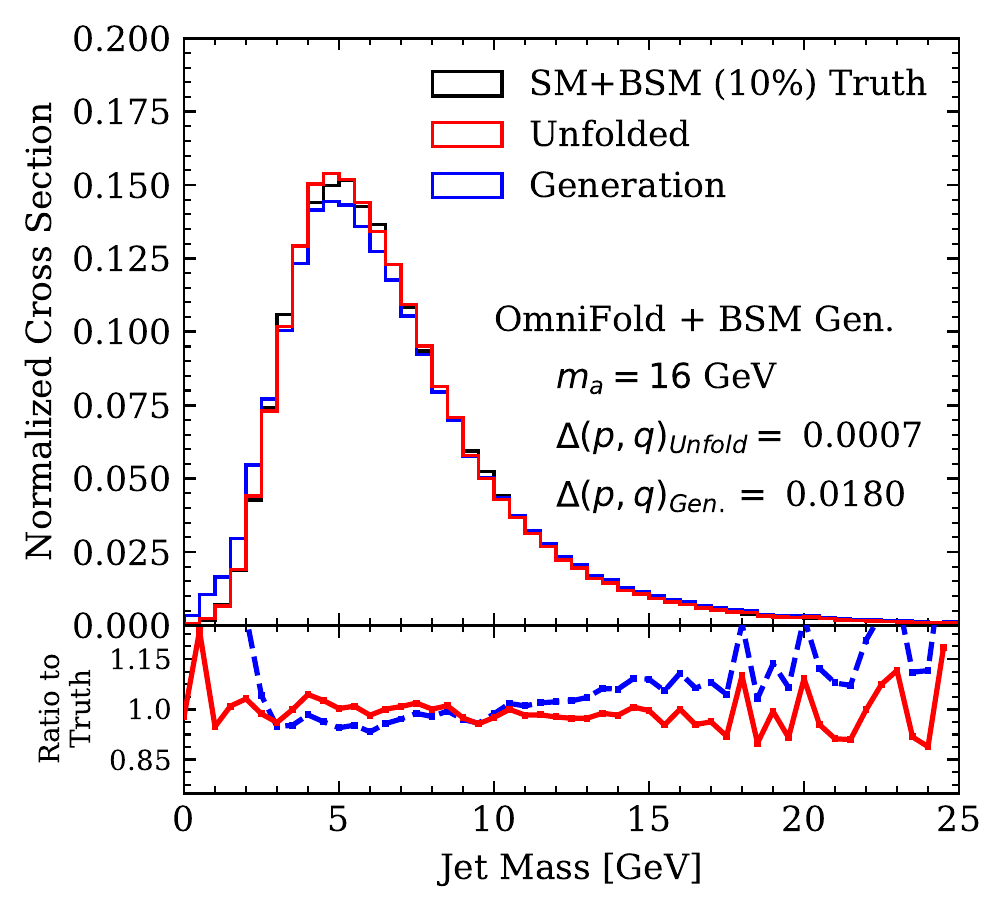}
  \includegraphics[height=3.5cm]{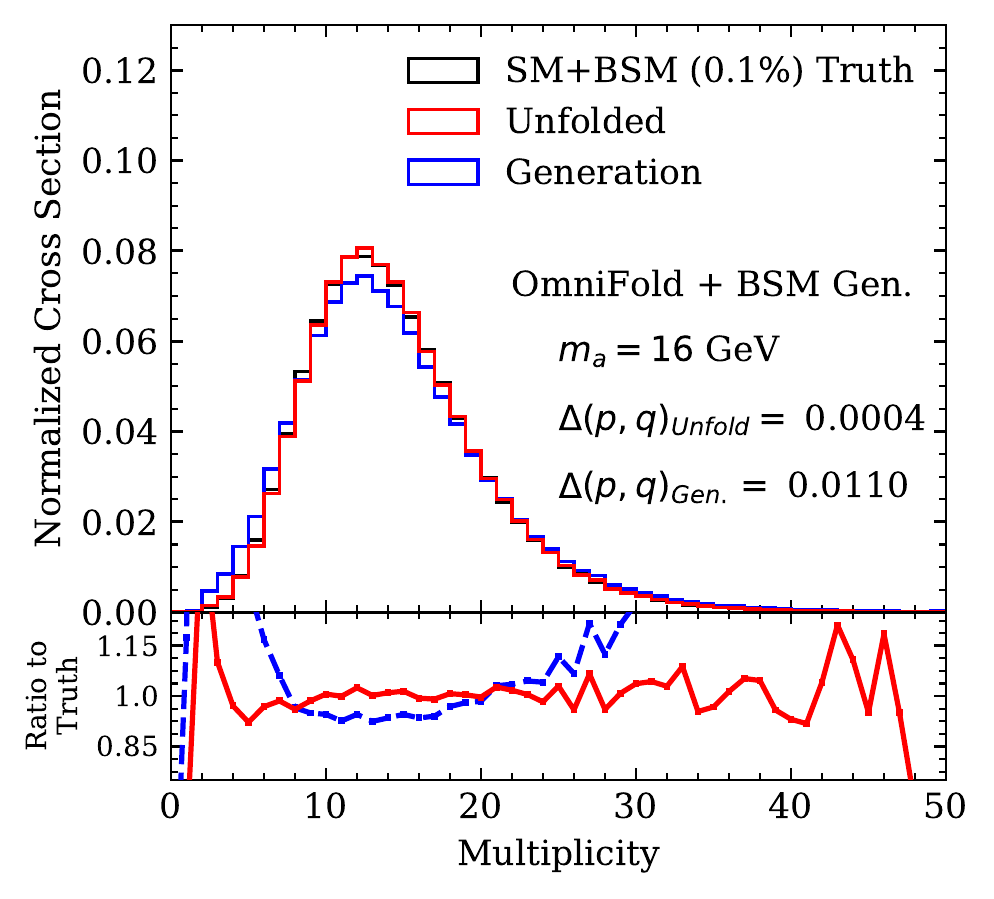}
  \includegraphics[height=3.5cm]{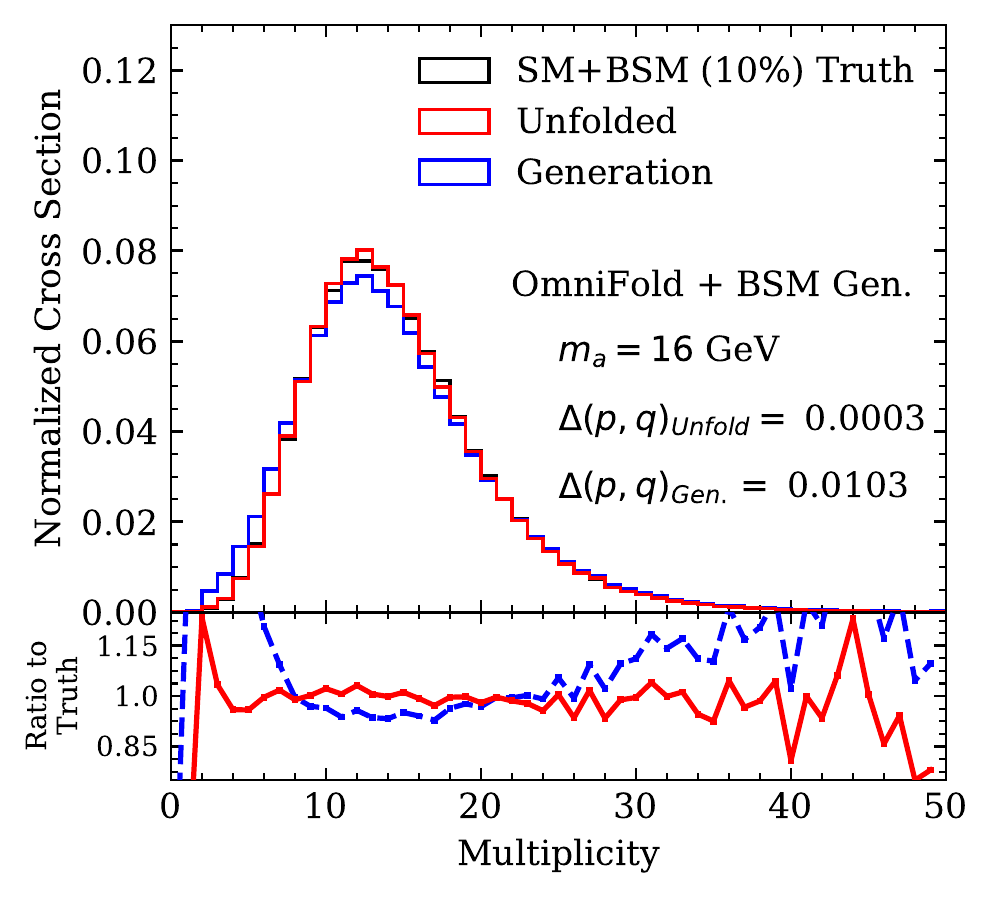}
  \caption{Truth, Generation, and unfolded Generation distributions for \omnifold, where BSM $h\rightarrow Za, a\rightarrow gg$ events have been included in the Truth \textit{and} the Generation.  Standard Model events in these samples come from \textsc{Pythia}~8 $Z$+jets simulation.  In the left column, 200 out of 200,000 Truth events come from the BSM sample, and in the right column, 20,000 out of 200,000 Truth events are BSM physics.  In all cases, 60,000 Generation events out of 260,000 come from $h\rightarrow Za, a\rightarrow gg$ events with different $m_a$ values.  Distributions are given for the invariant mass of the $Z$+jet, the jet mass, and the jet multiplicity.  The ratios of the Generation distributions are given to Truth for each plot.  The weights are taken after 5 iterations of \omnifold.}
  \label{fig:omni_16GeV_BSMGen}
\end{figure}

Similar to the end of Sec.~\ref{sec:Higgs250}, we explore how the performance in the previous section changes if we add in BSM to the Generation.  This can effectively populate the regions of phase space that are under-populated by the SM to enable a more precise post-unfolding search. 
For this purpose, we take the same 200,000 SM events in Sec.~\ref{sec:omnifold_125} and add 10,000 events each from $h\rightarrow Za, a\rightarrow gg$ samples with $m_a$ = 0.5, 1, 2, 4, 8, and 16 GeV, for a total of 260,000 events in Generation.  The \omnifold~method is carried out in exactly the same way as in Sec.~\ref{sec:omnifold_125}.  Distributions for the invariant $Z$+jet mass, jet mass, and jet multiplicity are shown in Fig.~\ref{fig:omni_16GeV_BSMGen}.  By comparing the triangular discriminator metric between Fig.~\ref{fig:omni_16GeV_BSMGen} and Fig.~\ref{fig:omni_16GeV}, it can be seen that when BSM physics is included in the Generation, the distributions generally agree slightly better after unfolding, particularly in the case of 10\% contamination, despite worse initial agreement.
\\

\begin{figure}
  \centering
  \includegraphics[height=5.5cm]{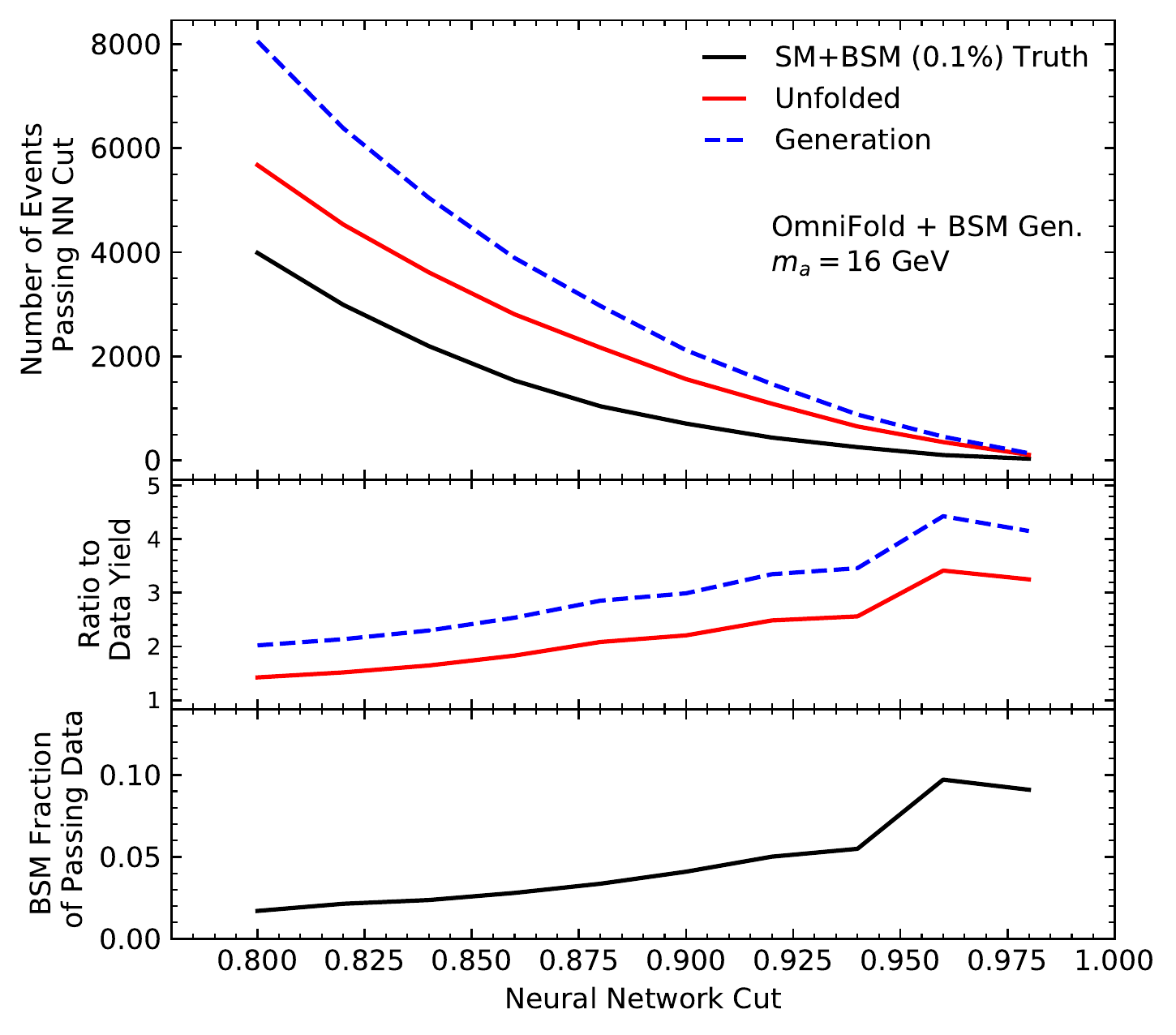}
  \includegraphics[height=5.5cm]{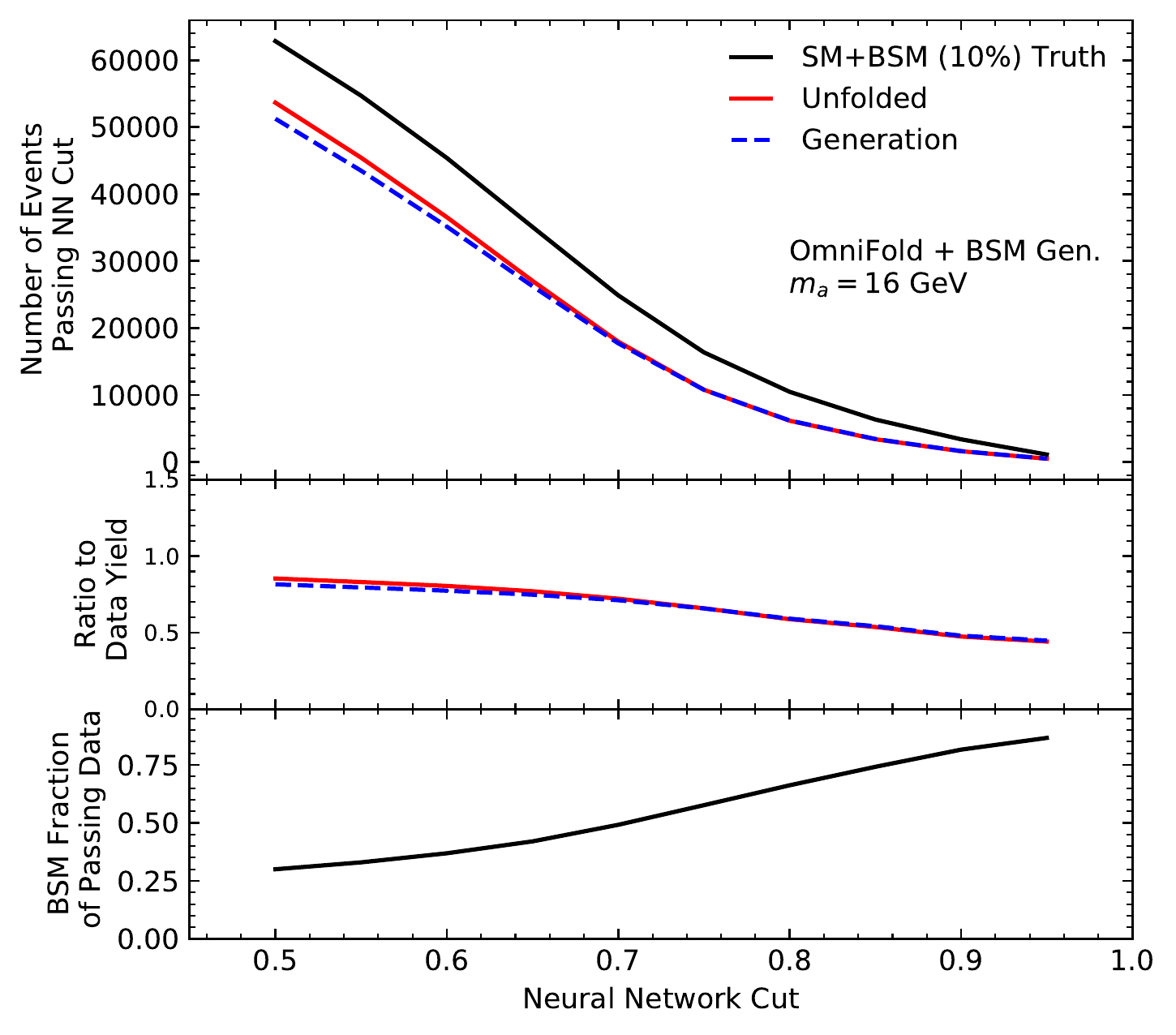}
  \caption{The number of Truth, Generation, and unfolded Generation events passing a cut on the PFN score, as a function of the cut value, in the case that 0.1\% of the data comes from BSM physics (top) and 10\% of the data comes from BSM physics (bottom).  In both cases, 60,000 Generation events out of 260,000 come from $h\rightarrow Za, a\rightarrow gg$ events with different $m_a$ values.  The unfolding was performed with \omnifold.  The PFN was specifically trained to distinguish SM from BSM physics.  The middle segment of both plots shows the ratio of the pre- and post-unfolding Generation yields to the Truth yield.  The bottom segment shows the fraction of Truth events that are BSM events at truth-level.}
  \label{fig:omnifold_discriminator_BSMGen}
\end{figure}

The SM vs. BSM discriminator PFN of Sec.~\ref{sec:omnifold_125} can applied to this new Generation sample.  The results of this application are shown in Fig.~\ref{fig:omnifold_discriminator_BSMGen}.  The PFN is also able to discriminate events with different $m_a$ values relatively accurately\footnote{Such that few events with $m_a = 4$ GeV will pass the NN cut, for example}, and it is clear here that the \omnifold~reweighted sample does not predict post-cut yields well.  The average weights found for the different $m_a$ components of Generation are given in Table~\ref{tab:weights_BSMGen}.  While the $m_a$ = 16 GeV events \textit{are} upweighted relative to the lighter $m_a$ events, it is clear that in the 0.1\% contamination case, the $m_a = 16$ GeV events are not adequately downweighted, and in the 10\% contamination case, they are not adequately upweighted.  Together with Fig.~\ref{fig:omnifold_discriminator}, it can be seen that while \omnifold~is performed using the full phase space, it has difficulty properly weighting extreme regions of phase space that can be particularly useful to model-dependent searches.

\begin{table}[htb]
\begin{center}
\begin{tabular}{ c|c|c }
Event Type & \multicolumn{1}{p{2cm}}{\centering Avg. Weight \\ (0.1\% Case)} & \multicolumn{1}{|p{2cm}}{\centering Avg. Weight \\ (10\% Case)} \\ \hline
$m_a$ = 0.5 GeV & 0.47 & 0.50 \\
$m_a$ = 1 GeV & 0.51 & 0.54 \\
$m_a$ = 2 GeV & 0.63 & 0.64 \\
$m_a$ = 4 GeV & 0.76 & 0.78 \\
$m_a$ = 8 GeV & 0.75 & 0.80 \\
$m_a$ = 16 GeV & 0.75 & 0.81 \\
SM & 0.81 & 0.80 \\
\end{tabular}
\caption{Average weights applied to events in Generation based on the event type.  Here Generation was 260,000 events, with 200,000 SM $Z$+jet events, and 10,000 events each from $h\rightarrow Za, a\rightarrow gg$ samples with different $m_a$ values, as given in the table.  Generation was used as an initial distribution for unfolding to a Truth sample of 200,000 events that either had 0.1\% or 10\% of its events drawn from the $h\rightarrow Za, a\rightarrow gg$ sample with $m_a$ = 16 GeV.  The average weight is 0.77 to match the normalization of the Truth sample, which has 200,000 events total, rather than 260,000.}
\label{tab:weights_BSMGen}
\end{center}
\end{table}

\section{Conclusions and Outlook}
\label{sec:conclusions}

The \omnifold~and \multifold~methods can be used for unbinned, all-variable unfolding in the presence of BSM physics, but there are inherent limitations on its applicability for truth-level searches for new physics.

In general, the distributions of high-level observables are unfolded well, as in Fig.~\ref{fig:Higgs250}, ~\ref{fig:multi_16GeV}, and~\ref{fig:omni_16GeV}.  This would enable model-independent searches or searches that use relatively high-level variables as discriminants, especially if the new physics has a high cross-section.  However, it is possible to devise strong BSM vs. SM discriminating variables that are not necessarily unfolded well, such as the neural network scores shown in Fig.~\ref{fig:multifold_discriminator} and~\ref{fig:omnifold_discriminator}.  These discriminants, which probe relatively subtle regions of phase space would most likely be applied in a model-dependent search.  While \omnifold~uses the full phase space, it has difficulty unfolding such specialized variables.  The best performance highlighted above is the 10\% BSM contamination case with \multifold, where the post-cut yield in the unfolded sample closely matches that found in Data.  In the 0.1\% contamination case with \multifold, there is also an enhancement in the reweighted sample relative to the Generation sample, but the agreement with Truth is not as stable as the 10\% case.  Together with the lack of agreement in the \omnifold~case, this suggests that it would be difficult to make a discovery of BSM physics unless the new physics comprises $>$ 1\% of data events.  Such high rates of BSM contamination would likely be discovered through conventional means by experimentalists prior to the release of unfolded datasets.

A significant issue in any attempt to perform a search with unfolded data is the inverse problem highlighted by Fig.~\ref{fig:Higgs250}.  Information is lost as particles pass through the detector, as seen in the smearing of the truth-level peak.  We have shown how this can be partially recovered by adding BSM events to the Generation.  This also helps to populate the most BSM-like regions of phase space.  For example, Fig.~\ref{fig:Higgs250_BSMGen} shows that this can be a powerful means to accurately reproduce an invariant mass peak even at truth-level.  However, this raises the natural question of how to choose the correct events to include in Generation.  The study in Sec.~\ref{sec:BSMGen} highlights the fact that even though high-level distributions can be unfolded well when BSM events are included in Generation, specialized variables may not be unfolded well; in particular, the reweighted distributions in Fig.~\ref{fig:omni_16GeV_BSMGen} are significantly different from both the Data \textit{and} from what would be expected in a SM-only case.  Because of this, a model-dependent search with the PFN discriminator would be ineffective in the 10\% case and return a false positive in the 0.1\% case.

Overall, our studies have shown that full phase space unfolding is a promising direction for post-measurement searches for resonant new physics.  However, significant work is required to increase the precision of the unfolding and to cope with cases where there are phase space regions with a large likelihood ratio.  It is likely that non-resonant new physics, which may be modeled using effective field theory methods, will be more successful because the likelihood ratio is never too far from unity.  This is closer to the previously studied case that investigated the impact of different SM simulations~\cite{Andreassen:2019cjw}.  The resonant examples presented in this paper will serve as an important benchmark for the community as existing methods are extended and new techniques are developed to empower a new class of analyses at the LHC and beyond.


\section*{Code and Data}

The code for this paper can be found at \url{https://github.com/wpmccormack/OmniFoldBSM}.

\begin{acknowledgments}
We thank Eric Metodiev for collaboration in the early stages of this project.  BN and WPM are supported by the U.S. Department of Energy (DOE), Office of Science under contract DE-AC02-05CH11231.  PK is supported by the National Science Foundation under Cooperative Agreement PHY-2019786 (The NSF AI Institute for Artificial Intelligence and Fundamental Interactions, \url{http://iaifi.org/}), and by the U.S. DOE Office of High Energy Physics under grant number DE-SC0012567.  We would also like to thank Jesse Thaler for useful discussions.
\end{acknowledgments}

\bibliography{HEPML,myrefs}

\end{document}